\documentclass[manuscript]{acmart}
\usepackage{amsmath}

\usepackage{amssymb}
\usepackage{amsthm}
\usepackage{mathtools}
\usepackage{algorithm}
\usepackage{algpseudocode}
\usepackage{algorithmicx}
\usepackage{graphicx}
\graphicspath{{figures/}}
\usepackage{float}
\usepackage{placeins}
\usepackage{array}
\usepackage{textcomp}
\usepackage{tabularx}
\usepackage{adjustbox}
\usepackage{multirow}
\usepackage{booktabs}
\usepackage{makecell}
\usepackage{hhline}
\usepackage{longtable}
\usepackage{rotating}
\usepackage{lscape}
\usepackage{enumitem}
\usepackage{threeparttable}
\usepackage{xurl}
\usepackage{subcaption}
\usepackage{tablefootnote}
\usepackage{pifont}
\usepackage[table]{xcolor}
\usepackage{colortbl}
\usepackage{listings}
\usepackage{tcolorbox}
\tcbuselibrary{listings,skins,breakable}

\usepackage{etc}
\usepackage{url}
\hypersetup{
  colorlinks=true,
  linkcolor=black,
  citecolor=black,
  urlcolor=black,
  pdftitle={Hierarchical Fault Detection and Diagnosis for Transformer Architectures},
  pdfauthor={Sigma Jahan, Saurabhsingh Rajput, Tushar Sharma, Mohammad Masudur Rahman},
  pdfsubject={Software engineering for deep learning; transformer fault diagnosis},
  pdfkeywords={transformer, fault diagnosis, attention, mutation testing, deep learning, software engineering}
}
\usepackage[capitalize]{cleveref}

\crefname{equation}{Eq.}{Eqs.}
\Crefname{equation}{Eq.}{Eqs.}

\definecolor{codebg}{HTML}{FAFAFA}
\definecolor{codeframe}{HTML}{CCCCCC}
\definecolor{codekw}{HTML}{006699}
\definecolor{codecm}{HTML}{3F7F5F}
\definecolor{codestr}{HTML}{7A3E9D}

\lstdefinestyle{friendly}{
  basicstyle=\ttfamily\scriptsize,
  keywordstyle=\color{codekw}\bfseries,
  commentstyle=\color{codecm},
  stringstyle=\color{codestr},
  showstringspaces=false,
  breaklines=true,
  breakatwhitespace=false,
  tabsize=2,
  columns=fullflexible,
  keepspaces=true
}

\tcbset{
  thesis-code/.style={
    listing engine  = listings,
    colback         = codebg,
    colframe        = codeframe,
    boxrule         = 0.4pt,
    arc             = 0pt,
    left            = 2.4em,
    right           = 4pt,
    top             = 5pt,
    bottom          = 5pt,
    breakable,
    enhanced,
    listing only,
  },
  thesis-code-scriptsize/.style={
    thesis-code,
    listing options = {style=friendly, language=Python, numbers=left, numbersep=10pt,
      breaklines=true, breakatwhitespace=false},
  },
}

\usepackage{newfloat}
\DeclareFloatingEnvironment[
  fileext=lop,
  listname={List of Listings},
  name=Listing,
  placement=tbp,
  within=section,
]{pylisting}
\crefname{pylisting}{Listing}{Listings}
\Crefname{pylisting}{Listing}{Listings}

\makeatletter
\AtBeginEnvironment{pylisting}{\let\@vspace\@vspace@orig\let\@vspacer\@vspacer@orig}
\makeatother

\usepackage{xspace}
\newcommand{\ie}{i.e.,\xspace}
\newcommand{\eg}{e.g.,\xspace}

\definecolor{markgreen}{HTML}{2F8F4E}
\definecolor{markred}{HTML}{C0392B}
\definecolor{oursrow}{RGB}{234,240,250}
\newcommand{\xmark}{\textcolor{markred}{\ding{55}}}
\newcommand{\cmark}{\textcolor{markgreen}{\ding{51}}}

\definecolor{dppband}{HTML}{ECEEF3}
\definecolor{dppbandtext}{HTML}{1C2A3A}
\definecolor{dpprule}{HTML}{2A2F36}
\definecolor{dppstBcol}{HTML}{2F6F3F}
\definecolor{dppstEUcol}{HTML}{1F4E89}
\definecolor{dppstELcol}{HTML}{8A4A1F}
\newcommand{\stB}{\textbf{\textsf{\textcolor{dppstBcol}{B}}}}
\newcommand{\stEU}{\textbf{\textsf{\textcolor{dppstEUcol}{EU}}}}
\newcommand{\stEL}{\textbf{\textsf{\textcolor{dppstELcol}{EL}}}}
\newcommand{\opid}[1]{\texttt{\small #1}}

\definecolor{rqblue}{RGB}{23,43,77}
\definecolor{rqbluebg}{RGB}{242,246,252}
\newtcolorbox{rqsummary}{
    colback=rqbluebg,
    colframe=rqblue,
    boxrule=0.5pt,
    arc=1.5mm,
    left=6pt, right=6pt, top=6pt, bottom=6pt,
    breakable
}
\newlist{rqlist}{itemize}{1}
\setlist[rqlist]{label=\small$\blacktriangleright$, leftmargin=1.5em, itemsep=6pt, parsep=4pt}

\newcounter{findingctr}

\usepackage{tikz}
\usetikzlibrary{fit,backgrounds,positioning,calc,arrows.meta,decorations.pathreplacing,shapes.geometric,shapes.misc}

\newcommand{\sbullet}{\(\scalebox{0.7}{$\bullet$}\)}

\theoremstyle{plain}

\theoremstyle{definition}

\theoremstyle{remark}

\crefname{definition}{Definition}{Definitions}
\Crefname{definition}{Definition}{Definitions}

\begin{document}
\title{Hierarchical Fault Detection and Diagnosis for Transformer Architectures}

\author{Sigma Jahan}
\affiliation{%
  \institution{Dalhousie University}
  \city{Halifax}
  \country{Canada}}
\email{sigma.jahan@dal.ca}

\author{Saurabhsingh Rajput}
\affiliation{%
  \institution{Dalhousie University}
  \city{Halifax}
  \country{Canada}}
\email{saurabh@dal.ca}

\author{Tushar Sharma}
\affiliation{%
  \institution{Dalhousie University}
  \city{Halifax}
  \country{Canada}}
\email{tushar@dal.ca}

\author{Mohammad Masudur Rahman}
\affiliation{%
  \institution{Dalhousie University}
  \city{Halifax}
  \country{Canada}}
\email{masud.rahman@dal.ca}

\renewcommand{\shortauthors}{Jahan et al.}
\settopmatter{printacmref=false}
\begin{abstract}
Transformers now underpin critical AI systems across industry and research. Yet their faults can silently alter model behavior without runtime errors, and existing techniques offer little support for tracing these failures to their component and root cause. Such faults evade detection because loss and numerical values stay normal, and the visible symptom rarely identifies the component responsible. We present \emph{DEFault++}, a hierarchical learning-based technique that first detects a fault, then identifies the affected component, and finally the cause within it, helping developers effectively debug transformer models. DEFault++ organizes component-level runtime measurements with a \emph{Fault Propagation Graph (FPG)}, a structural prior over the architecture's dependency paths, and reports the evidence behind each diagnosis. To train and evaluate it, we construct \emph{DEFault-bench}, a benchmark of 5{,}556 labeled runs from mutation testing across seven models, nine tasks, and both encoder and decoder architectures. DEFault++ improves fault detection over four prior techniques, reaching an F1 of 0.826--0.909, and in a developer study with 21 participants, it raises repair accuracy from 57.1\% to 83.3\%. These results show that transformer fault diagnosis benefits from component-level measurements and architecture-aware reasoning rather than model-level behavior alone, and DEFault-bench provides a foundation for further research on transformer fault diagnosis.

\end{abstract}



\maketitle

\providecommand{\symbf}[1]{\boldsymbol{#1}}
\section{Introduction}
\label{sec:defaultpp_introduction}
Transformer models now support a wide range of applications, from natural language processing~\citep{gpt4,gemini} and code generation~\citep{copilot} to broader software engineering tasks~\citep{softwaredevelopment} and medical imaging~\citep{medicalimaging}. Their effectiveness comes in large part from the attention mechanism~\citep{vaswani2017attention}, which lets each token draw on information from any other token when forming its representation. Attention combines several operations~\citep{vaswani2017attention}, such as projection, masking, and score computation, and these operations interact with other transformer components such as residual connections and layer normalization, as well as with runtime settings such as kernel selection and numerical precision~\citep{attentiondetails}. Small mistakes in these operations or their interactions can change model behavior without raising explicit runtime errors~\citep{ABNN_sigma,attentionPlease,zhai2023stabilizing}. The model can lose representation quality or task performance even when training completes without exceptions or invalid values.

A fault can affect a model's behavior without producing any of the usual error symptoms. More than a third of attention-specific faults produce \emph{silent} or latent symptoms, about twice the rate reported for traditional deep neural network (DNN) faults~\citep{ABNN_sigma}. These faults survive conventional tests, reach production, and waste training resources precisely because they raise no error. Even when a fault becomes visible, the symptom rarely indicates which component is responsible, which leaves a developer without a clear starting point for debugging or repair. Weak fine-tuning performance, for example, can come from a faulty attention mask, stale key-value cache entries, or incorrect positional information~\citep{ABNN_sigma}. Diagnosing a transformer fault therefore means identifying both the faulty component and the cause that explains its behavior.

Existing work has proposed automated debugging techniques for DNN programs. Rule-based techniques, such as AutoTrainer~\citep{autotrainer}, UMLAUT~\citep{umlaut}, and DeepDiagnosis~\citep{deepdiagnosis}, check training runs for predefined symptoms such as exploding gradients, vanishing gradients, or oscillating loss. Learning-based techniques, such as DeepFD~\citep{deepfd} and DEFault~\citep{default}, use runtime features to predict DNN fault categories such as loss-function, learning-rate, or activation faults. Both lines of work remain useful for architecture-agnostic model and training faults, but they observe behavior at the model level and do not localize faults to transformer-specific components such as masking, the query-key-value (QKV) projection, positional encoding, or key-value caching. Transformer-specific debuggers inspect attention behavior more directly. ATTNChecker~\citep{liang2025attnchecker} detects \texttt{NaN} or \texttt{Inf} values during attention computation, and AtPatch~\citep{weng2026atpatch} localizes attention-map anomalies at the patch level, but each targets a narrow failure mode rather than the full range of transformer faults. As a result, current techniques report only coarse, model-level categories such as incorrect activation or incorrect loss, without identifying which transformer component caused a fault or which cause inside that component explains it. Silent faults are the hardest case for this model-level view, since they leave the loss curves and finite numerical values these techniques monitor unchanged.

To address this gap, we propose DEFault++, a hierarchical learning-based technique for diagnosing faults in transformer models. DEFault++ produces three levels of diagnosis. It first detects whether a fault is present, then assigns the fault to a transformer component category such as the QKV projection or masking, and finally identifies the root cause within that category. Together, these three levels point a developer toward the responsible component and the cause behind the observed behavior rather than toward a generic training symptom, so a developer knows which component to inspect and why it is faulty. Even fault detection is difficult here, since a fault can shift internal attention behavior while the overall training metrics look normal. Our key idea is that each such fault still leaves a distinctive pattern inside the affected component. DEFault++ measures how each transformer component behaves during training and connects these measurements through a structural prior, the \emph{Fault Propagation Graph (FPG)}, derived from the transformer's forward and backward equations. This prior organizes the runtime measurements so the diagnostic model can link an observed deviation to a fault category and report the measurements behind each diagnosis. Training and evaluating DEFault++ requires transformer faults labeled at the component level, which no existing benchmark provides. We therefore construct DEFault-bench, a benchmark of 5{,}556 labeled instances from encoder and decoder architectures, generated with a transformer-specific mutation technique. On DEFault-bench, DEFault++ improves fault detection over four prior DNN debugging techniques, reaching an F1 of up to 0.91 for fault detection. To confirm that these findings extend to faults found in practice, we further evaluate DEFault++ on real-world faults collected from open-source repositories. In a developer study with 21 participants, repair-action accuracy rises from 57.1\% without assistance to 83.3\% with DEFault++.

In summary, we make the following contributions:
\begin{enumerate}[label=(\alph*), topsep=3pt]
\item We propose \emph{DEFault++}, a hierarchical technique that detects a fault in a transformer model, categorizes the affected component, and identifies the root cause within that component.
\item We construct \emph{DEFault-bench}, a benchmark of 5{,}556 labeled transformer instances across encoder and decoder architectures, and release the transformer-specific mutation and fault-injection framework behind it, which others can reuse to generate further transformer fault data~\citep{defaultpp_repo}.
\item We define a \emph{Fault Propagation Graph (FPG)}, a structural prior that encodes the dependency paths among transformer components and gives the diagnostic model the transformer's architectural structure.
\item We design a training objective for root-cause diagnosis that combines supervised contrastive learning with prototype matching, together with an FPG-based explanation that reports which feature groups support each diagnosis.
\item We evaluate DEFault++ on DEFault-bench, against four prior techniques, on real-world faults, and in a developer study with 21 participants.
\end{enumerate}

\section{Motivating Example}
\label{sec:defaultpp_motivation}

To show the kind of fault DEFault++ targets, consider the real-world fault in \cref{lst:qkv_fusion_bug}, taken from the Hugging Face Diffusers library (Issue~\#11903\footnote{\url{https://github.com/huggingface/diffusers/issues/11903}}). The method \texttt{fuse\_qkv\_projections} merges the separate query, key, and value projections into one fused layer and routes the forward pass through that fused layer. The original projection modules stay in the model so that LoRA~\citep{hu2022lora} can still attach adapters to them. The forward pass, however, no longer uses those modules, so LoRA updates parameters that no longer affect the model output. The run completes without exceptions, the loss decreases, and no invalid values appear. The only visible sign of the fault is that fine-tuning does not change model behavior.

\begin{pylisting}[!ht]
\tcbinputlisting{
    thesis-code-scriptsize,
    unbreakable,
    listing file={qkv_fusion_bug.py},
}
\caption{A real-world transformer fault (Hugging Face Diffusers Issue~\#11903).}
\label{lst:qkv_fusion_bug}
\end{pylisting}

Diagnosing this fault is hard for existing techniques. A static analysis that inspects the model structure or parameter list still finds valid \texttt{q\_proj}, \texttt{k\_proj}, and \texttt{v\_proj} modules, so it cannot tell that their updates no longer reach the forward pass. Rule-based dynamic analyses that monitor training, such as AutoTrainer~\citep{autotrainer}, UMLAUT~\citep{umlaut}, and DeepDiagnosis~\citep{deepdiagnosis}, do not flag the run because it shows none of their predefined symptoms, such as \texttt{NaN} or \texttt{Inf} values, exploding gradients, or oscillating loss. Learned classifiers, such as DeepFD~\citep{deepfd} and DEFault~\citep{default}, recognize predefined DNN fault categories such as activation-function faults, so they offer little help for a fault in a transformer QKV projection path. Transformer debuggers such as ATTNChecker~\citep{liang2025attnchecker} and AtPatch~\citep{weng2026atpatch} inspect attention behavior more directly, but they target narrower symptoms such as invalid attention scores, which this example does not produce. In contrast, DEFault++ measures runtime traces such as QKV alignment, attention entropy, and projection update activity during training. These measurements characterize the faulty QKV projection path, so DEFault++ detects the fault, assigns it to the QKV category, and identifies the root cause as a stale projection update path.

\section{Background and Related Work}
\textbf{Fault Taxonomies.} Empirical studies of deep learning (DL) programs classify faults by root cause, covering faults in training procedures, hyperparameters, and layer configurations across feed-forward, convolutional, and recurrent architectures~\citep{faulttaxonomy,dlbugcharacterstics}. These taxonomies supply the fault categories used by several DL diagnosis techniques, but they predate attention-based architectures and do not describe faults that arise from attention computation, masking, projection paths, positional encodings, key-value caching, or transformer-specific kernel behavior. In our prior work, we addressed part of this gap with a taxonomy of attention faults that classifies 555 attention faults from open-source projects into seven categories and 25 root causes~\citep{ABNN_sigma,attention_taxonomy}. The present work builds its fault labels on these attention-specific categories together with five non-attention categories from the DL fault taxonomies~\citep{faulttaxonomy,dlbugcharacterstics}, which cover the surrounding model and training faults such as loss, learning-rate, and activation faults (\cref{sec:fault-taxonomy}).\\

\noindent \textbf{Mutation Testing for Deep Learning.}
Mutation testing assesses how well a test suite detects faults by introducing small, controlled changes into a program, called \emph{mutants}, and checking whether the tests reveal them~\citep{rlmutation}. The deep learning community has adapted this idea to neural models. DeepMutation~\citep{deepmutation} and DeepMutation++~\citep{deepmutation++} introduce mutation operators that modify model structure, training code, or learned parameters, and report mutation scores to quantify test adequacy. DeepCrime~\citep{deepcrime} injects faults into the source code at training time, reruns training several times, and applies statistical tests with effect-size thresholds to decide whether a mutation changes behavior. This repetition matters because the same injected change can look harmful under one random seed and negligible under another~\citep{mosbach2021stability}. PyTorchFI~\citep{pytorchfi} injects faults at the tensor and bit levels to study the fault tolerance of PyTorch models. Mutation testing can also supply labeled training data for fault diagnosis. DeepFD~\citep{deepfd} injects five common training faults, such as an incorrect learning rate, to create labeled samples for supervised diagnosis, and later work extends this idea to larger and more diverse sets of faulty DNN programs~\citep{deep4deep, default, rlmutation}. These datasets mostly cover architecture-agnostic DNN faults, such as loss misconfiguration, incorrect activation functions, or incorrect layer counts. They do not target the transformer-specific mechanisms or attention-related behavior needed to diagnose transformer faults (\cref{tab:related-comparison}).

\begin{table}[htbp]
\caption{Comparison of the DEFault-bench mutation process with prior mutation techniques. \cmark{} marks a supported dimension and \xmark{} one that is not supported.}
\label{tab:related-comparison}
\footnotesize
\renewcommand{\arraystretch}{1.0}
\setlength{\tabcolsep}{4pt}
\begin{tabularx}{\linewidth}{@{} l *{4}{>{\centering\arraybackslash}X} >{\columncolor{oursrow}\centering\arraybackslash}X @{}}
\toprule
\textbf{Dimension} & \textbf{DeepMut.++~\citep{deepmutation++}} & \textbf{DeepCrime~\citep{deepcrime}} & \textbf{DeepFD~\citep{deepfd}} & \textbf{PyTorchFI~\citep{pytorchfi}} & \textbf{DEFault-bench} \\
\midrule
Fault granularity        & Layer/param & Source (AST) & Program & Bit/tensor & Transformer unit \\
Mechanism count          & 17 & 24 of 35 & 5 types & Bit-flip & 12 categories \\
Architecture coverage    & FFNN/CNN/RNN & Generic DNN & Generic DNN & CNN (infer.) & Encoder + decoder \\
Attention mechanisms     & \xmark & \xmark & \xmark & \xmark & \cmark \\
Behavior change          & \xmark & \xmark & \xmark & \xmark & \cmark \\
Decision basis           & Accuracy & GLM + Cohen's $d$ & Accuracy & SDC rate & Accuracy + sign-flip \\
Noise control            & \xmark & \cmark & \xmark & \xmark & \cmark \\
Scale (reported)         & Varies & 1{,}760 & 52 programs & $10^7$+ inj. & 5{,}556 \\
Labeled traces           & \xmark & \xmark & \cmark & \xmark & \cmark \\
\bottomrule
\end{tabularx}
\end{table}

\noindent
\textbf{Debugging Techniques for Deep Learning Programs.}
Existing DL debugging techniques fall into three broad groups. The first group learns fault patterns from training behavior. DeepFD~\citep{deepfd} extracts four feature types, namely loss, gradient, weight, and activation features, and classifies five common training faults. Our prior work DEFault~\citep{default} extends this idea into a hierarchical classifier that first detects whether a program is faulty and then assigns it to one of seven generic DNN fault categories. These two methods are the closest dynamic baselines to our work, since they infer fault labels from training-time measurements, and DEFault++ extends our hierarchical design from generic DNN fault categories to transformer-specific components. They observe behavior at the model level and report model-level fault categories. As a result, they cannot tell whether a symptom comes from the QKV projection, masking, positional encoding, residual connections, normalization, key-value caching, or another transformer component. DEFault++ targets exactly this finer, component-level granularity.

\begin{table}[htbp]
\caption{Comparison of fault diagnosis techniques for DNN programs. FD, FC, and RC denote fault detection, fault categorization, and root-cause diagnosis, and a parenthetical count gives the number of categories or root causes a technique reports. DEFault++ covers 45 root causes in total, 40 for encoders and all 45 for decoders. \cmark{} marks a supported capability and \xmark{} one that is not supported.}
\label{tab:related_work_comparison}
\footnotesize
\setlength{\tabcolsep}{10pt}
\renewcommand{\arraystretch}{1.0}
\begin{tabular}{@{} l c c c c c l @{}}
\toprule
\textbf{Technique} & \textbf{Target} & \textbf{FD} & \textbf{FC} & \textbf{RC} & \textbf{Attn-Aware} & \textbf{Explanation} \\
\midrule
AutoTrainer~\citep{autotrainer}     & DNN         & \cmark              & Partial (5)      & \xmark    & \xmark    & Rule-based          \\
UMLAUT~\citep{umlaut}               & DNN         & \cmark              & Partial          & \xmark    & \xmark    & Rule-based          \\
DeepDiagnosis~\citep{deepdiagnosis} & DNN         & \cmark              & \cmark~(8)       & \xmark    & \xmark    & Decision tree       \\
DeepLocalize~\citep{deeplocalize}   & DNN         & NaN/Inf only        & \xmark           & \xmark    & \xmark    & Numerical trace     \\
DeepFD~\citep{deepfd}               & DNN         & \cmark              & \cmark~(5)       & \xmark    & \xmark    & Feature attribution \\
DEFault~\citep{default}             & DNN         & \cmark              & \cmark~(7)       & \xmark    & \xmark    & SHAP                \\
ATTNChecker~\citep{liang2025attnchecker} & Transformer & NaN/Inf only   & \xmark           & \xmark    & \cmark    & \xmark              \\
AtPatch~\citep{weng2026atpatch}         & Transformer & Map outlier         & \xmark           & \xmark    & \cmark    & \xmark              \\
FT-Transformer~\citep{fttransformer} & Transformer & Soft errors        & \xmark           & \xmark    & \cmark    & \xmark              \\
\midrule
\rowcolor{oursrow}
\textbf{DEFault++ (Ours)} & Transformer & \cmark & \cmark~(12) & \cmark~(45) & \cmark & FPG+prototype \\
\bottomrule
\end{tabular}
\end{table}

Rule-based and heuristic techniques monitor predefined failure symptoms. AutoTrainer~\citep{autotrainer}, UMLAUT~\citep{umlaut}, and DeepDiagnosis~\citep{deepdiagnosis} check training-time symptoms such as exploding gradients, vanishing gradients, and oscillating loss. DeepLocalize~\citep{deeplocalize} and GRIST~\citep{GRIST} focus on numerical instability, while Tensfa~\citep{tensfa} and TFCheck~\citep{ben2023testing} check tensor shapes and training-process invariants. Other methods work at finer granularity, including MODE~\citep{mode}, DeepFault~\citep{deepfault}, Apricot~\citep{apricot}, and DeepSeer~\citep{deepseer}. These techniques work well when a fault produces one of the symptoms or structural patterns they monitor. They have limited reach, however, for transformer faults that keep loss curves normal, tensor shapes valid, and numerical values finite while changing internal attention behavior.

A third group of techniques analyzes a model statically, before any training run, so it operates at a different level of abstraction from ours. NeuraLint~\citep{NeuraLint}, DEBAR~\citep{debar}, and NerdBug~\citep{nerdbug} check meta-model graphs, tensor abstractions, and API usage, while DeepCheck~\citep{gopinath2019symbolic} and CRADLE~\citep{cradle} analyze trained networks or compare backend implementations. FL4Deep~\citep{fl4deep} sits between the static and dynamic views, combining both through a knowledge graph to rank candidate root causes. These techniques find structural, symbolic, API-level, or backend-level faults, but they do not diagnose transformer-component faults from training traces. We therefore treat them as complementary to DEFault++ rather than as direct baselines.\\

\noindent
\textbf{Transformer-Specific Fault Analysis.} A smaller body of work analyzes transformer-internal behavior, but each technique targets a narrower failure mode than the full transformer fault taxonomy (\cref{fig:defaultpp-fault-taxonomy}). ATTNChecker~\citep{liang2025attnchecker} detects NaN or Inf values during attention computation, AtPatch~\citep{weng2026atpatch} localizes attention-map anomalies at the patch level, FT-Transformer~\citep{fttransformer} studies hardware soft-error tolerance for vision transformers, and Bug Attention Probe~\citep{bap2025} applies attention probing to code-understanding tasks. None of these techniques produces a combined hierarchy of fault detection, component category, and root-cause diagnosis. A related direction perturbs inputs to study model behavior~\citep{jones2020adversarial,ribeiro2020checklist,clark2019does,voita2019analyzing,deeptest}. Their goal differs from ours, since these methods expose behavioral failures through modified inputs, whereas we diagnose implementation faults that arise during ordinary training. Debuggers that use large language models (LLMs), such as SoapFL~\citep{soapfl2025}, LLM4FL~\citep{llm4fl2024}, ChatDBG~\citep{chatdbg2025}, and BugReAct~\citep{islam2026agents}, reason over source code, execution logs, or bug reports to localize faults in general software. They are also complementary, since they operate after a developer provides software artifacts or failure reports, while DEFault++ reads training-time traces from transformer components to infer a transformer-specific fault category and root cause.\\

\noindent
\textbf{Explainability for Fault Diagnosis.}
Diagnostic output is more useful when it gives developers evidence for a prediction rather than only a label, because the evidence shows where to look and gives a reason to trust the diagnosis. Existing explanation methods fall into two families. Post-hoc methods compute explanations after a trained model makes a decision. SHAP~\citep{lundberg2017unified}, DiCE~\citep{mothilal2020dice}, and Anchors~\citep{ribeiro2018anchors} return feature attributions, counterfactual examples, or rule-based explanations, and our prior method DEFault adopts this post-hoc approach~\citep{default}. Inherently interpretable methods instead build explanation into the prediction mechanism. Prototype networks~\citep{snell2017prototypical} explain a prediction through similarity to class prototypes, and concept-bottleneck models~\citep{koh2020concept} route predictions through human-readable intermediate concepts. DEFault++ follows this interpretable-by-design approach and reports the training-time evidence behind each transformer fault diagnosis.

\section{Study Design}
\label{sec:defaultpp_study_design}
Our study comprises three parts. We first construct DEFault-bench, a transformer fault benchmark built through mutation testing that provides labeled examples of correct and faulty fine-tuning behavior. We then design DEFault++, a hierarchical diagnostic technique that detects a fault, categorizes the affected component, and identifies its root cause, organized around how faults propagate across transformer components. Finally, we evaluate DEFault++ on DEFault-bench, on real-world faults, and through a developer study. We organize the evaluation around four research questions that move from whether the technique works, to how it compares with prior work, to which of its design choices matter, to whether the evidence it reports can be trusted.

\begin{rqlist}
\item \textbf{RQ$\mathbf{_1}$ (Effectiveness):} How effectively does DEFault++ detect, categorize, and diagnose the root cause of transformer faults? No prior technique resolves a transformer fault to both its component and its root cause, so the first step is to establish whether the three-level design works at all. This question measures each level on the evaluated transformer models, before any comparison with prior work.
\item \textbf{RQ$\mathbf{_2}$ (Comparison):} How does DEFault++ compare with existing deep learning fault-detection techniques? Prior techniques target generic DNN faults and observe behavior at the model level, so it is unclear whether they detect faults that surface only inside transformer components. This question tests whether component-level measurement adds detection value over these model-level techniques.
\item \textbf{RQ$\mathbf{_3}$ (Design contribution):} How much do the Fault Propagation Graph and the root-cause separation training contribute to diagnostic performance? DEFault++ combines several design choices, and a strong overall result does not show which of them carry the performance. This question isolates the contribution of the propagation structure and the separation training at each diagnosis level.
\item \textbf{RQ$\mathbf{_4}$ (Explanation faithfulness):} Do the runtime feature groups that DEFault++ reports as evidence drive its root-cause diagnosis? A diagnosis is more useful when it comes with evidence a developer can inspect, but reported evidence is trustworthy only if it reflects the model's actual decision. This question tests whether the reported feature groups, rather than incidental ones, determine the predicted root cause.
\end{rqlist}

\section{Experimental Setup}
\label{sec:defaultpp_setup}
This section describes the transformer models and tasks we study and the cross-validation protocol we use to evaluate generalization. The benchmark construction (\cref{sec:defaultpp_dataset}) and the diagnostic technique (\cref{sec:defaultpp_approach}) both build on this setup.

\subsection{Subject Models and Tasks}
\label{sec:defaultpp_experimental_setup}
We select the transformer models for our study by three criteria. A model must be widely used, openly available through a standard Hugging Face Transformers implementation~\citep{huggingface_transformers}, and small enough, at most 125M parameters, to fine-tune repeatedly under matched seeds within our compute budget. Applying these criteria gives four encoder-only and three decoder-only models that together cover both architecture families, two depths, and distilled and full-size variants of the same family. The encoder models are BERT-base (110M parameters, 12 layers)~\citep{devlin2019bert}, RoBERTa-base (125M, 12)~\citep{liu2019roberta}, DistilBERT (66M, 6)~\citep{sanh2019distilbert}, and DistilRoBERTa (82M, 6)~\citep{sanh2019distilbert}. The decoder models are GPT-2 (124M, 12)~\citep{radford2019language}, GPT-Neo-125M (125M, 12)~\citep{black2021gptneo}, and DistilGPT-2 (82M, 6)~\citep{sanh2019distilbert}.

We select downstream tasks that are commonly adopted in prior transformer and deep learning studies and that come with openly released datasets. Each task is a publicly available benchmark with an established evaluation metric, and the set spans both classification and language modeling so that the injected faults are exercised under different training objectives. For the encoder models, we use five GLUE classification tasks~\citep{wang2018glue}, namely SST-2, QNLI, RTE, MRPC, and QQP, with accuracy as the task metric. For the decoder models, we use four language-modeling corpora, namely LAMBADA~\citep{paperno2016lambada}, PTB~\citep{marcus1993ptb}, WikiText-2~\citep{merity2017wikitext}, and OpenWebText~\citep{gokaslan2019openwebtext}, with log-perplexity as the task metric. All models use their default tokenizers and the standard transformer-block structure that our mutation operators target.

We fine-tune every model with a single shared setup across tasks, namely AdamW~\citep{loshchilov2019adamw} with linear warmup and mixed precision~\citep{devlin2019bert,mosbach2021stability}, so that observed differences trace to the injected fault rather than to hyperparameter variation. We repeat each fine-tuning run under five fixed seeds $\mathcal{S} = \{42, 123, 456, 789, 101112\}$~\citep{mosbach2021stability} and vary the affected layers and the fault severity uniformly, so that no single layer or magnitude dominates the benchmark.

\subsection{Cross-Validation Protocol}
\label{sec:defaultpp_cv}
We evaluate generalization to unseen model--task pairs using nested grouped cross-validation with $k=5$ outer folds. The grouping unit is the model--task pair, so all runs from the same pair remain in the same fold. This prevents correlated traces from the same model and task from appearing in both training and test data. The encoder subset contains 20 model--task pairs (4 models $\times$ 5 GLUE tasks), and the decoder subset contains 12 model--task pairs (3 models $\times$ 4 language-modeling tasks). Because we repeat each run across seeds, we average those per-seed measurements into one example, so the random seed is not used as a grouping variable. The outer loop evaluates held-out model--task pairs, and the inner loop uses \texttt{StratifiedGroupKFold} for model selection~\citep{wardhani2019cross}. We fit preprocessing steps and tune hyperparameters only inside the inner training fold, and all methods use the same outer-fold assignments. Because each outer fold holds out whole model--task pairs, a fold-level standard deviation would mostly reflect which pairs are held out rather than training noise, so we report results at the model--task level.

\section{DEFault-bench: Benchmark Construction}
\label{sec:defaultpp_dataset}
\begin{figure}[htbp]
    \centering
    \includegraphics[width=0.75\textwidth]{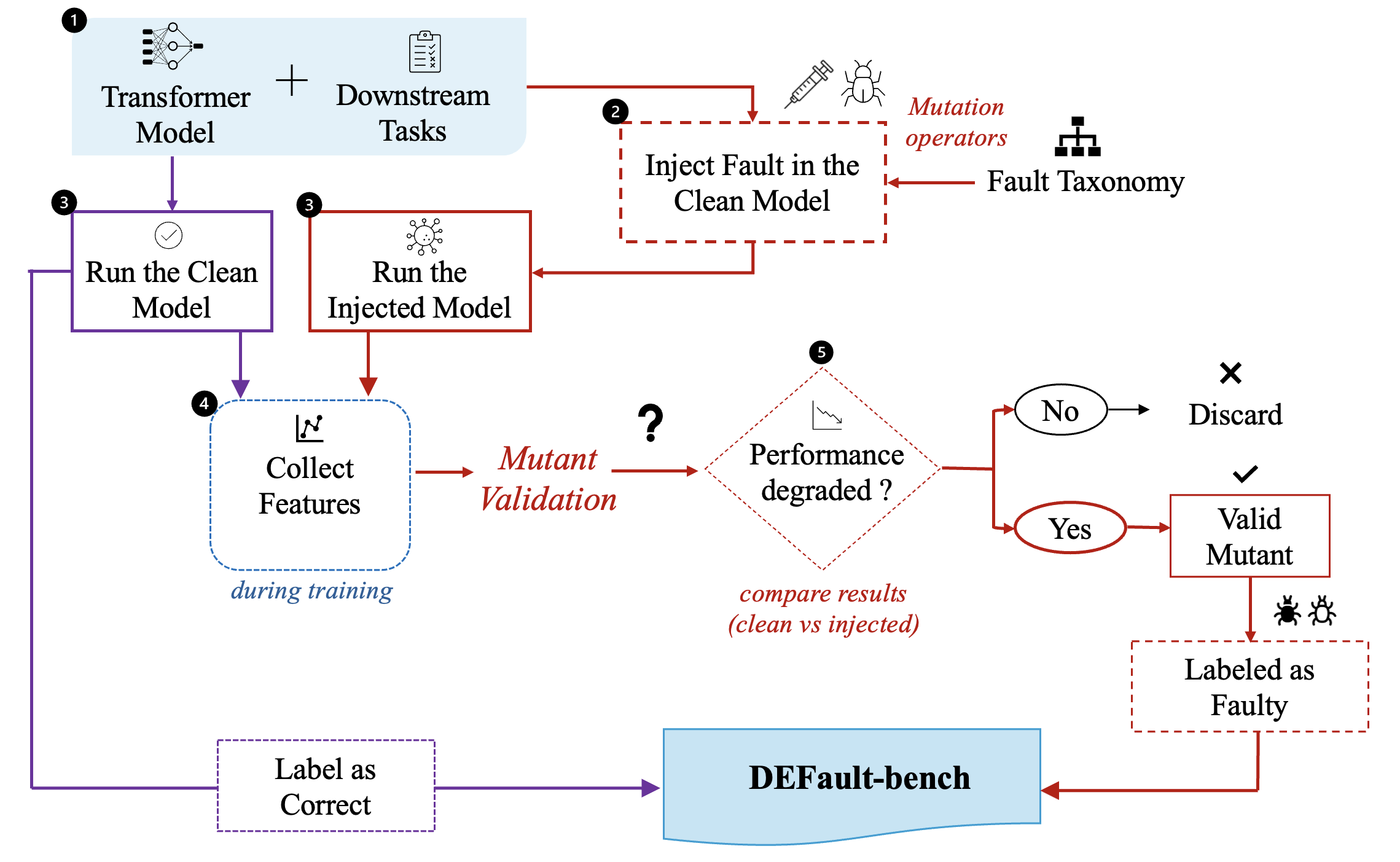}
    \Description{Workflow diagram for constructing DEFault-bench}
    \caption{Workflow for constructing DEFault-bench}
    \label{fig:dpp-dataset-workflow}
\end{figure}

Training and evaluating a component-level diagnostic technique requires transformer faults labeled by component, which no existing benchmark provides. We therefore construct DEFault-bench, a benchmark of clean and faulty fine-tuning runs for the seven transformer models and nine tasks introduced in \cref{sec:defaultpp_setup}. We create the faulty runs with a transformer-specific mutation technique that injects documented attention and DNN faults into the components of the transformer architecture, either by changing stored parameters or by changing computations during the forward pass. We organize these faults with a fault taxonomy (\cref{sec:fault-taxonomy}) and realize each one as a mutation operator (\cref{sec:mutation-operator}). Each injected fault is one configuration $\mathcal{C} = (m, t, u, f, v, \ell, \sigma)$ that fixes the model, task, target component, fault category, mutation variant, target layer, and severity. We validate each mutant in two steps. We first check that the intended structural change was applied. We then test whether the faulty run behaves differently from its matched clean run under the same seeds, following prior mutation-testing work~\citep{deepcrime,deepmutation} (\cref{fig:dpp-dataset-workflow}).

\subsection{Fault Taxonomy}
\label{sec:fault-taxonomy}
\begin{figure}[htbp]
    \centering
    \includegraphics[width=0.80\textwidth]{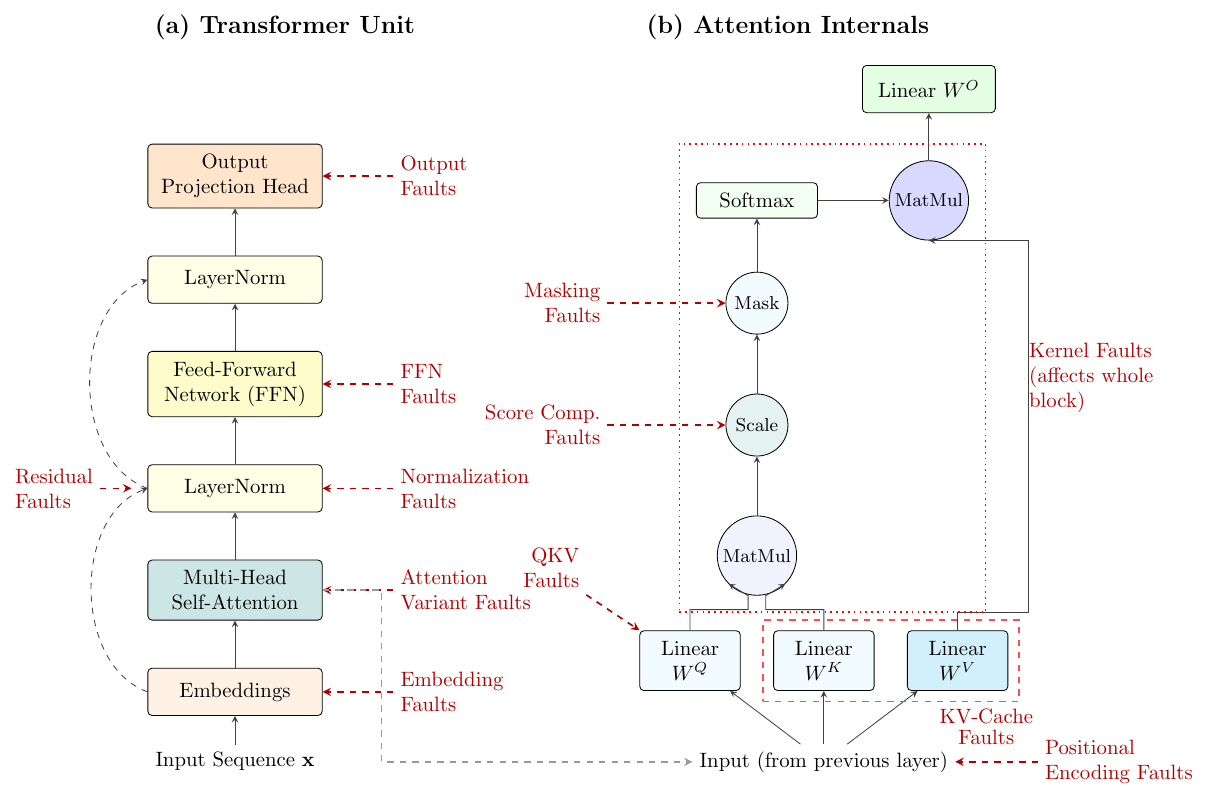}
    \caption[Fault categories organized by transformer components.]{Fault categories organized by transformer components}
    \Description{Diagram showing fault categories organized by transformer component, with block-level and attention-internal categories.}
     \label{fig:transformer-block}
\end{figure}

We base the fault taxonomy on the main components of a transformer architecture (\cref{fig:transformer-block}). A transformer model contains input embeddings, a stack of repeated transformer blocks, and an output projection head~\citep{vaswani2017attention}. Each block contains multi-head self-attention (MHA), a position-wise feed-forward network (FFN), residual connections, and layer normalization (LayerNorm). For an input sequence $\mathbf{x} \in \mathbb{R}^{n \times d}$, a post-norm block adds each sublayer's output back to its input and then normalizes the sum, computing $\mathbf{h} = \mathrm{LayerNorm}(\mathbf{x} + \mathrm{MHA}(\mathbf{x}))$ for the attention sublayer and $\mathbf{y} = \mathrm{LayerNorm}(\mathbf{h} + \mathrm{FFN}(\mathbf{h}))$ for the feed-forward sublayer. We group faults by six components, namely embeddings, MHA, FFN, layer normalization, residual connections, and the output projection head. The same component set applies to pre-norm variants, where only the position of LayerNorm changes.

\begin{figure}[htbp]
\centering
\includegraphics[width=\textwidth]{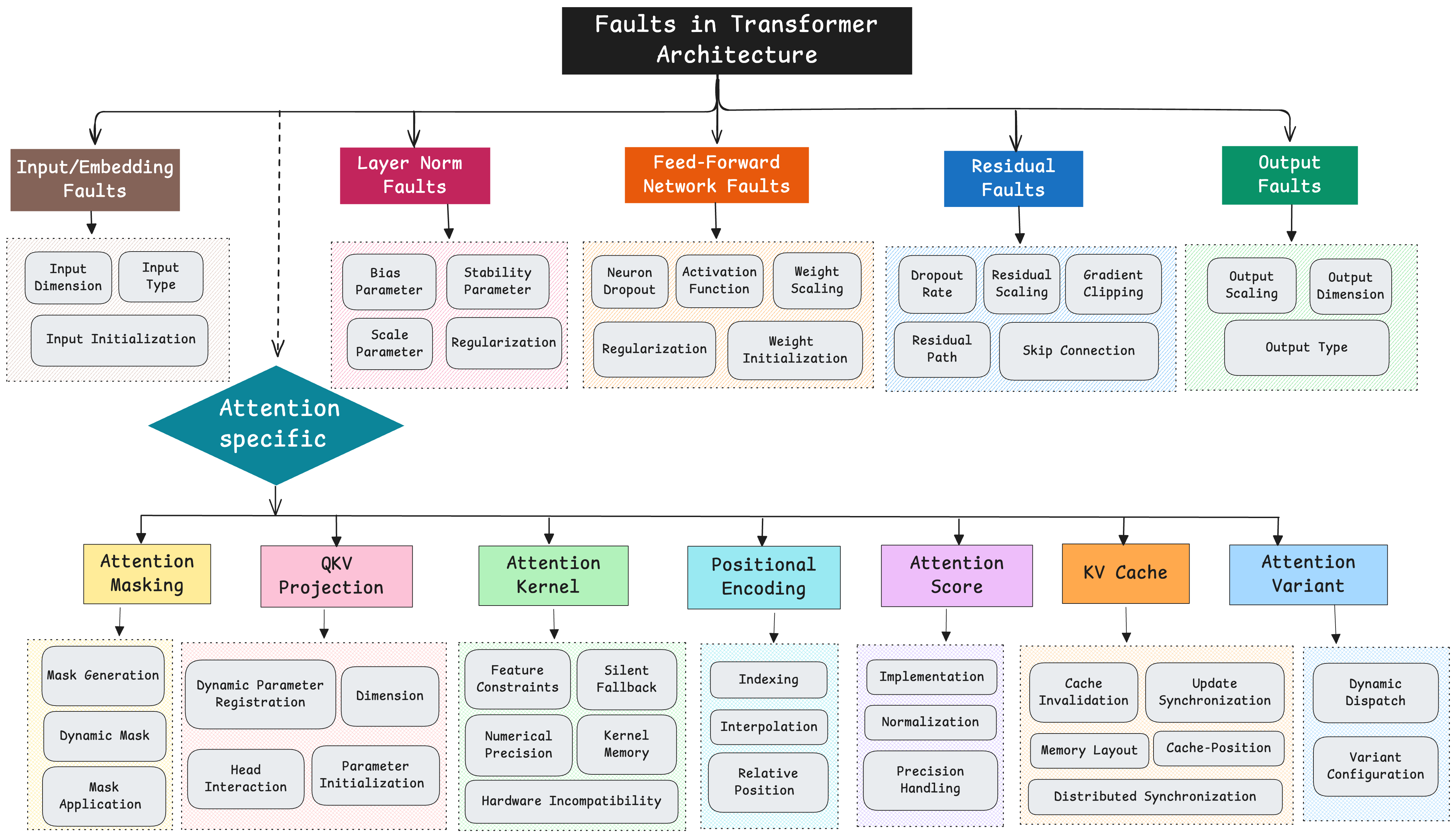}
\caption{Taxonomy of transformer fault categories and root-cause labels used by DEFault++}
\Description{A taxonomy diagram showing transformer fault categories and root-cause labels used by DEFault++.}
\label{fig:defaultpp-fault-taxonomy}
\end{figure}

We then map these transformer components to documented fault categories and root causes (\cref{fig:defaultpp-fault-taxonomy}). The attention-related categories come from our prior fault taxonomy for attention-based neural networks (ABNNs), which we derived from 555 real-world attention faults~\citep{ABNN_sigma,attention_taxonomy}. The non-attention categories come from two existing DNN fault taxonomies based on 970 faults in total~\citep{faulttaxonomy,dlbugcharacterstics}.

The seven attention categories map to different parts of the attention computation~\citep{ABNN_sigma, vaswani2017attention}, illustrated in the attention-internal view of \cref{fig:transformer-block},
\begin{equation}
\mathrm{Attention}(Q, K, V) = \mathrm{softmax}\!\left(\frac{QK^\top}{\sqrt{d_k}} + M\right)V
\label{eq:dpp-attention}
\end{equation}
where $Q$, $K$, and $V$ are linear projections of the input, $d_k$ is the key dimension, and $M$ is an additive mask. \emph{Masking} faults corrupt the mask $M$ through padding masks, causal masks, or mask reshaping across batch and head dimensions. \emph{QKV}, \emph{Score}, and \emph{Positional} faults affect the projections, the attention-score computation, and the positional information, respectively. \emph{Kernel} faults involve the configured attention backend or numerical execution path. \emph{Variant} faults use the wrong form of attention, such as single-head attention or causal masking in an encoder. \emph{KV Cache} faults apply to autoregressive decoders and affect cache-state management~\citep{adnan2024keyformer}. The remaining components each contribute one category, following prior DNN fault taxonomies~\citep{faulttaxonomy,dlbugcharacterstics}. \emph{Embedding} faults involve token, segment, or feature-dimension embeddings. \emph{FFN} faults cover feed-forward weights, neurons, and activation functions. \emph{LayerNorm} faults involve the learned scale $\gamma$, bias $\beta$, numerical-stability term $\epsilon$, or regularization of LayerNorm parameters~\citep{xiong2020layer,mosbach2021stability}. \emph{Residual} faults cover skip connections and residual dropout, and \emph{Output} faults involve the prediction head or logits.

Overall, the taxonomy contains 12 fault categories and 45 root causes, including five KV Cache root causes that apply only to autoregressive decoders. Transformer models without autoregressive cache behavior, such as encoder-only models, use 11 of these categories and 40 root causes (\cref{fig:defaultpp-fault-taxonomy}).

{\arrayrulecolor{dpprule}%
\begin{table}[htbp]
\centering
\scriptsize
\renewcommand{\arraystretch}{1.00}
\setlength{\tabcolsep}{3pt}
\caption{Attention-specific mutation operators used to construct DEFault-bench}
\label{tab:dpp-mutation-operators-attention}
\begin{tabularx}{\linewidth}{@{}
  >{\raggedright\arraybackslash}p{0.62in}
  >{\raggedright\arraybackslash}p{0.95in}
  >{\raggedright\arraybackslash}p{1.05in}
  >{\centering\arraybackslash}p{0.32in}
  >{\raggedright\arraybackslash}X
  >{\centering\arraybackslash}p{0.20in}
@{}}
\toprule
\textbf{Component} & \textbf{Root cause} & \textbf{Operator} & \textbf{ID} & \textbf{Executable Mutation Parameters} & \textbf{ST} \\
\midrule

\multirow{4}{=}{Masking}
 & Mask application & Zero attention mask         & \opid{MZM} & Replace valid attention mask with an all-zero mask of the same shape.                & \stB  \\
 & Mask application & Invert attention mask       & \opid{MIM} & Invert mask semantics while preserving expected mask shape \& dtype.                & \stB  \\
 & Mask generation  & Reshape mask incorrectly    & \opid{MRM} & \emph{axis}: batch or head dimension to misalign; retained only if broadcast-compatible. & \stEL \\
 & Dynamic mask     & Causal-mask break (decoder) & \opid{MCB} & \emph{visibility}: fraction of future keys made visible within valid mask tensor.    & \stEU \\
\cmidrule(l{2pt}r{2pt}){1-6}

\multirow{7}{=}{QKV Projection}
 & Parameter initialization       & Zero $Q$ projection         & \opid{QZQ} & Zero the existing Q projection weights without changing tensor shape.               & \stB  \\
 & Parameter initialization       & Zero $K$ projection         & \opid{QZK} & Zero the existing K projection weights without changing tensor shape.                 & \stB  \\
 & Parameter initialization       & Zero $V$ projection         & \opid{QZV} & Zero the existing V projection weights without changing tensor shape.               & \stB  \\
 & Head interaction               & Swap $Q \leftrightarrow K$  & \opid{QSW} & Swap compatible Q and K projection tensors or outputs.                            & \stB  \\
 & Head interaction               & Tie head weights            & \opid{QTH} & \emph{heads}: subset of existing heads tied while preserving projection dimensionality. & \stEL \\
 & Dynamic parameter registration & Freeze QKV gradients        & \opid{QFG} & Disable updates to existing QKV parameters while preserving the forward path.           & \stB  \\
 & Dimension                      & Repartition head dimensions & \opid{QHD} & \emph{split}: alternative $(\mathrm{heads}, d_{\mathrm{head}})$ head factorization with $\mathrm{heads}\cdot d_{\mathrm{head}}=d_{\mathrm{model}}$, retained only when broadcast-compatible. & \stEL \\
\cmidrule(l{2pt}r{2pt}){1-6}

\multirow{3}{=}{Score}
 & Normalization      & Drop $1/\!\sqrt{d_k}$ scaling & \opid{SDS} & Remove score scaling while preserving attention-score shape.                              & \stB  \\
 & Implementation     & Apply dropout pre-softmax     & \opid{SPD} & \emph{p}: dropout probability applied to score tensors before softmax.                    & \stEU \\
 & Precision handling & Unsafe fp16 cast              & \opid{SUC} & Cast score computation to fp16; configurations producing non-finite traces are discarded. & \stB  \\
\cmidrule(l{2pt}r{2pt}){1-6}

\multirow{3}{=}{Positional}
 & Indexing          & Omit positional embeddings  & \opid{POE} & Remove or zero the positional contribution while preserving hidden-state shape.       & \stB  \\
 & Relative position & Shift position indices      & \opid{PSI} & \emph{$\Delta$}: index shift restricted to valid supported positions.                 & \stEU \\
 & Interpolation     & Truncate positional support & \opid{PTL} & \emph{cutoff}: maximum retained position index; out-of-range indices are discarded.   & \stEU \\
\cmidrule(l{2pt}r{2pt}){1-6}

\multirow{5}{=}{Kernel}
 & Silent fallback          & Force slow backend                    & \opid{KSB} & Force a valid non-optimized attention backend.                                                & \stB  \\
 & Feature constraints      & Mismatched dropout probability        & \opid{KMD} & \emph{$p_{\mathrm{cfg}}, p_{\mathrm{kern}}$}: valid training vs.\ kernel dropout settings.     & \stEU \\
 & Hardware incompatibility & Trigger valid fallback (dtype/layout) & \opid{KFT} & \emph{dtype} or \emph{layout} mismatch that triggers a supported fallback path.                & \stEL \\
 & Numerical precision      & Reduced-precision kernel math         & \opid{KRP} & \emph{precision}: reduced-precision accumulation mode for the attention kernel. Configurations producing non-finite traces are discarded. & \stEL \\
 & Kernel memory            & Memory-constrained kernel path        & \opid{KMC} & Force the attention kernel onto a reduced-memory chunked execution path. Configurations producing non-finite traces are discarded.       & \stB  \\
\cmidrule(l{2pt}r{2pt}){1-6}

\multirow{2}{=}{Variant}
 & Variant configuration & Single effective head attention & \opid{VSH} & Route, tie, or mask heads so that only one effective head remains while preserving output shape. & \stB  \\
 & Dynamic dispatch      & Causal mask in encoder          & \opid{VEC} & Apply a broadcast-compatible causal mask in an encoder attention layer.                     & \stB  \\
\cmidrule(l{2pt}r{2pt}){1-6}

\multirow{5}{=}{KV Cache (decoder)}
 & Cache invalidation          & Stale cache         & \opid{CST} & \emph{layers}: subset of layers using stale but shape-compatible cached $K, V$ states.       & \stEL \\
 & Update synchronization      & Desynchronized cache update & \opid{CDU} & Withhold the current decoding step's freshly computed $K, V$ from the cache so attention and prediction read desynchronized cached states. Cache tensor shape is preserved. & \stB  \\
 & Cache-position              & Off-by-one indexing & \opid{COB} & \emph{shift}: cache index offset restricted to valid cache positions.                         & \stEU \\
 & Memory layout               & Truncate cache      & \opid{CTR} & \emph{length}: maximum retained cache length while preserving the expected cache interface.   & \stEU \\
 & Distributed synchronization & Cross-request leak  & \opid{CLK} & Reuse compatible cached states across requests without changing cache tensor shape.           & \stB  \\

\bottomrule
\end{tabularx}
\end{table}
}

\subsection{Mutation Operators}
\label{sec:mutation-operator}

To generate faults systematically, we realize each root cause as one or more \emph{mutation operators}, where a mutation operator is a specific code change that injects that root cause into a transformer model. We read the root causes from the taxonomy in \cref{fig:defaultpp-fault-taxonomy} and define an operator for each distinct way a root cause can be realized as a code change, which yields 52 operators across the 12 fault categories. We present the operators in two tables that follow the two taxonomy sources. The 29 attention-specific operators come from our attention fault taxonomy and target masking, the QKV projection, score computation, positional information, the attention kernel, attention variants, and the decoder KV cache (\cref{tab:dpp-mutation-operators-attention}). The 23 architecture-level operators come from the DNN fault taxonomies and target embeddings, the FFN, layer normalization, residual connections, and the output head (\cref{tab:dpp-mutation-operators-arch}). This grouping reflects where a fault originates in the model, which is separate from the run-time mechanism used to inject it (\cref{sec:defaultpp_injection}).

We give each operator a three-letter identifier, following the naming style of prior DL mutation work~\citep{deepcrime,deepmutation}. The first letter identifies the transformer component, and the remaining two letters form a short mnemonic for the change. For example, \opid{QZQ} uses \textbf{Q} for the QKV component, \textbf{Z} for zeroing, and \textbf{Q} for the query projection, while \opid{SPD} uses \textbf{S} for the Score component, \textbf{P} for pre-softmax, and \textbf{D} for dropout. A fault category can hold more than one operator because one root cause can be triggered through more than one separate change. The QKV parameter-initialization root cause, for example, has three operators that zero the query, key, or value projection (\opid{QZQ}, \opid{QZK}, \opid{QZV}), since each projection is a separate entry point for the same root cause. This is why the 52 operators map to 45 root causes rather than one operator each.

The Search Type (ST) column records how an operator selects the parameter listed in the Executable Mutation Parameters column. A \stB{} (binary) operator takes no parameter and applies a single fixed change. A \stEU{} operator has a numeric parameter swept over a predefined list of values, such as a scaling factor or a dropout probability. A \stEL{} operator has a categorical parameter chosen from a fixed set, such as \opid{FCA}, which replaces the activation function with each function in $\{\mathrm{ReLU}, \mathrm{GELU}, \mathrm{Tanh}, \mathrm{Sigmoid}\}$. For parameterized operators, our mutation technique evaluates low, medium, and high settings, so the benchmark spans a range of fault strengths rather than only extreme mutants. Extreme mutants carry little diagnostic information, since almost any test exposes them~\citep{deepcrime}.

{\arrayrulecolor{dpprule}%
\begin{table}[htbp]
\centering
\scriptsize
\renewcommand{\arraystretch}{1.00}
\setlength{\tabcolsep}{3pt}
\caption{Architecture-level mutation operators used to construct DEFault-bench}
\label{tab:dpp-mutation-operators-arch}
\begin{tabularx}{\linewidth}{@{}
  >{\raggedright\arraybackslash}p{0.62in}
  >{\raggedright\arraybackslash}p{0.95in}
  >{\raggedright\arraybackslash}p{1.05in}
  >{\centering\arraybackslash}p{0.32in}
  >{\raggedright\arraybackslash}X
  >{\centering\arraybackslash}p{0.20in}
@{}}
\toprule
\textbf{Component} & \textbf{Root cause} & \textbf{Operator} & \textbf{ID} & \textbf{Executable Mutation Parameters} & \textbf{ST} \\
\midrule

\multirow{4}{=}{Embedding}
 & Input initialization & Zero token embedding subset    & \opid{ETZ} & \emph{percentage}: fraction of vocabulary entries zeroed while preserving embedding shape. & \stEU \\
 & Input type           & Swap embedding pairs           & \opid{ESW} & \emph{percentage}: fraction of token pairs swapped within existing vocabulary.         & \stEU \\
 & Input type           & Scale segment / type embedding & \opid{ESS} & \emph{factor}: multiplicative scale applied to existing segment or type embeddings.        & \stEU \\
 & Input dimension      & Zero embedding feature dimensions & \opid{EZD} & \emph{fraction}: contiguous block of embedding feature dimensions zeroed while embedding tensor shape is preserved. & \stEU \\
\cmidrule(l{2pt}r{2pt}){1-6}

\multirow{5}{=}{FFN}
 & Weight scaling        & Scale FFN weights            & \opid{FSW} & \emph{factor}: multiplicative scale on existing $W_1, W_2$ tensors.                       & \stEU \\
 & Neuron dropout        & Permanently drop neurons     & \opid{FDN} & \emph{percentage}: fraction of hidden neurons zeroed while preserving FFN shape.          & \stEU \\
 & Activation function   & Change activation function   & \opid{FCA} & \emph{activation}: executable replacement nonlinearity (\eg~GELU $\to$ ReLU).             & \stEL \\
 & Regularization        & Change weight regularization & \opid{FRG} & \emph{scheme}: replacement weight-decay or $L_2$ coefficient accepted by the optimizer.   & \stEU \\
 & Weight initialization & Change weight initialization & \opid{FWI} & \emph{init}: replacement initialization scheme applied to existing $W_1, W_2$ tensors.    & \stEL \\
\cmidrule(l{2pt}r{2pt}){1-6}

\multirow{5}{=}{LayerNorm}
 & Scale parameter     & Scale $\gamma$    & \opid{NSG} & \emph{factor}: multiplicative scale on existing $\gamma$ parameters.          & \stEU \\
 & Scale parameter     & Zero $\gamma$     & \opid{NZG} & Zero existing $\gamma$ parameters while preserving LayerNorm shape.           & \stB  \\
 & Bias parameter      & Shift $\beta$     & \opid{NSB} & \emph{shift}: additive offset on existing $\beta$ parameters.                  & \stEU \\
 & Stability parameter & Change $\epsilon$ & \opid{NCE} & \emph{value}: positive numerical-stability constant accepted by LayerNorm.   & \stEU \\
 & Regularization      & Weight-decay LayerNorm parameters & \opid{NWD} & \emph{coefficient}: weight-decay coefficient applied to the LayerNorm $\gamma, \beta$ parameters, which are normally excluded from decay. & \stEU \\
\cmidrule(l{2pt}r{2pt}){1-6}

\multirow{5}{=}{Residual}
 & Skip connection   & Remove skip connection   & \opid{RRS} & Zero/bypass residual branch while preserving sublayer output shape.            & \stB  \\
 & Residual scaling  & Scale residual branch    & \opid{RSR} & \emph{factor}: multiplicative scale on the residual path.                              & \stEU \\
 & Residual path     & Inject Gaussian noise    & \opid{RIN} & \emph{$\sigma$}: standard deviation of additive noise with matching tensor shape.     & \stEU \\
 & Gradient clipping & Change gradient clipping & \opid{RGC} & \emph{value}: replacement max-norm clip threshold accepted by the training loop.      & \stEU \\
 & Dropout rate      & Change residual dropout rate & \opid{RDR} & \emph{p}: residual-branch dropout probability accepted by the training loop.      & \stEU \\
\cmidrule(l{2pt}r{2pt}){1-6}

\multirow{4}{=}{Output}
 & Output scaling   & Scale output logits             & \opid{OSL} & \emph{factor}: multiplicative scale on logits while preserving output dimension.                                        & \stEU \\
 & Output dimension & Zero rows for selected classes  & \opid{OZR} & \emph{classes}: subset of existing output rows to zero.                                                                & \stEL \\
 & Output type      & Reinitialize output projection  & \opid{ORI} & \emph{init}: replacement initialization scheme applied to existing output projection.                              & \stEL \\
 & Output dimension & Change output interface         & \opid{OOD} & \emph{dim/map}: root-cause-derived output-dimension or mapping fault; retained only when compatible with the task loss.& \stEU \\

\bottomrule
\end{tabularx}
\end{table}
}

\subsection{Fault Injection Mechanism}
\label{sec:defaultpp_injection}
We inject every fault programmatically, so each configuration is applied, evaluated, and reverted without manual edits to the model source, and we release this injection framework in our replication package~\citep{defaultpp_repo}. We denote the clean model parameters by $\theta$ and the injected mutant by $\theta'$, and we evaluate the two under matched random seeds and otherwise identical fine-tuning conditions, so that any measured difference reflects the injected configuration $\mathcal{C}$. Recall that $\mathcal{C} = (m, t, u, f, v, \ell, \sigma)$ fixes the model architecture, the downstream task, the target unit, the fault category, the fault variant, the affected layer indices $\ell \subseteq \{1,\dots,L\}$ for a model with $L$ transformer blocks, and the severity level $\sigma \in \{\mathrm{low}, \mathrm{medium}, \mathrm{high}\}$. The severity sets the magnitude of the injected change for numeric variants, such as scaling factors and dropout rates, and maps to a discrete intensity for non-numeric variants, such as the visibility ratio in causal-mask breaking. We use only single-fault configurations, with one target unit, one category, and one variant per run, so that each observed change traces to one injected fault. Transformer faults can affect either stored parameters or forward-pass computations, so injection uses two mechanisms~\citep{pytorchfi,liang2025attnchecker}.

\emph{Static faults} change parameter tensors at rest, before forward execution. An FFN weight-scaling fault, for example, multiplies the targeted weight matrices by a scalar,
\begin{equation}
\theta'_i =
\begin{cases}
\alpha \cdot \theta_i & \text{if } i \in \mathrm{params}(u,\ell),\\[0.2ex]
\theta_i & \text{otherwise}
\end{cases}
\label{eq:dpp-static-fault}
\end{equation}
\noindent where $\mathrm{params}(u,\ell)$ selects the parameters of unit $u$ in layers $\ell$. Static faults include embedding corruption, parameter scaling, permanent neuron dropout, and LayerNorm parameter changes. \emph{Dynamic faults} change the forward computation path through wrappers or hooks at selected module call sites. A mask-zeroing fault, for example, intercepts the attention call and replaces the mask $M$ with zeros,
\begin{equation}
\texttt{forward}'(x, M) = \texttt{forward}(x, \mathbf{0}_{|M|})
\label{eq:dpp-dynamic-fault}
\end{equation}
\noindent where $\mathbf{0}_{|M|}$ is a zero tensor of the same shape as $M$. Dynamic faults include mask manipulation, $Q$/$K$ swapping, removal of score scaling, and forced kernel selection. In both mechanisms the transformer architecture stays unchanged, and only the targeted computation or parameter values change. \cref{fig:dpp-injection-code-panels} shows one representative example of each mechanism.

\begin{figure}[htbp]
\centering
\begin{minipage}[t]{0.48\textwidth}
\begin{tcblisting}{
  listing engine=listings,
  listing options={style=friendly, language=Python},
  colback=codebg, colframe=red!40, boxrule=0.4pt, arc=0pt,
  left=4pt, right=4pt, top=4pt, bottom=4pt,
  listing only, equal height group=rowDppA,
  title={\scriptsize\bfseries (a) Static fault injection (encoder FFN)}}
# static fault: mutate FFN weights at rest
layer = model.bert.encoder.layer[layer_idx]
ffn = layer
orig_w1 = ffn.intermediate.dense.weight
orig_w2 = ffn.output.dense.weight
self._backup = (orig_w1.clone(), orig_w2.clone())
with torch.no_grad():
    orig_w1.mul_(self.alpha)  # scale W1
    orig_w2.mul_(self.alpha)  # scale W2
\end{tcblisting}
\end{minipage}\hfill
\begin{minipage}[t]{0.48\textwidth}
\begin{tcblisting}{
  listing engine=listings,
  listing options={style=friendly, language=Python},
  colback=codebg, colframe=red!40, boxrule=0.4pt, arc=0pt,
  left=4pt, right=4pt, top=4pt, bottom=4pt,
  listing only, equal height group=rowDppA,
  title={\scriptsize\bfseries (b) Dynamic fault injection (decoder causal mask)}}
# dynamic fault: wrap the same model's attention.forward to weaken causal mask
self._backup_fwd = attention.forward
def faulty_forward(*args, **kwargs):
    mask = _get_attention_mask(kwargs)
    if mask is not None:
        # unmask a fraction of future keys
        m = _weaken_causal_mask(mask, visibility_ratio)
        kwargs["attention_mask"] = m
    return self._backup_fwd(*args, **kwargs)
attention.forward = faulty_forward
\end{tcblisting}
\end{minipage}
\caption[Injection mechanisms.]{Two injection mechanisms on the same model: (a) static parameter mutation scales FFN weights, (b) dynamic forward wrapping weakens the causal mask at runtime. The clean run leaves all modules unmodified.}
\Description{Two code panels showing the two injection mechanisms: static parameter mutation of FFN weights and dynamic wrapping of the attention forward pass to weaken the causal mask.}
\label{fig:dpp-injection-code-panels}
\end{figure}

Masking faults need special handling for decoder-only models, where self-attention is causal by construction and a token at position $i$ cannot attend to a future position $j>i$. The causal mask enforces this by placing $-\infty$ in the future positions of the additive mask $M$, so that softmax assigns those positions zero weight. To inject a causal violation, we weaken this constraint rather than add a new mask. We replace a configurable fraction of those $-\infty$ entries in the upper-triangular region of $M$ with $0$, controlled by the severity $\sigma$, so each query can attend to a proportion of keys beyond its own position. We treat causal violations as decoder instances of attention masking faults.

To keep clean and faulty runs comparable, we hold the main experimental settings fixed between the $\theta$ and $\theta'$ runs, such as the dataset, optimizer, and number of epochs, and we pair each faulty run with a clean run under the same seed. A context manager carries out the injection. It stores the original parameter tensors and forward methods, applies the fault specified by $\mathcal{C}$, and restores the original state after evaluation. Restoring the state at the end of each configuration prevents one injected fault from carrying over into the next run.

Finally, we apply a structural verification check to each injected fault. For static faults, we verify that the parameter difference is restricted to $\mathrm{params}(u,\ell)$ and that its magnitude matches $\sigma$ within a relative tolerance of $10^{-6}$ to account for floating-point rounding. For dynamic faults, we verify that hooks or wrappers attach only to the intended module call sites and that the original forward function is restored after the context manager exits. We also verify execution completeness by requiring each configuration to produce training logs, and we exclude configurations that crash or raise runtime errors.

\subsection{Mutant Validation}
\label{sec:defaultpp_mutant_validation}

Injecting a fault does not guarantee that it changes model behavior, so we decide which mutants to keep through a statistical mutation-killing test, following prior work~\citep{deepcrime}. For each configuration, \cref{alg:dpp-injection} injects the fault, verifies the structural change, fine-tunes the clean and mutant models under the shared seeds, and applies the killing test to the paired results.

\begin{algorithm}[htbp]
\caption{Per-configuration fault injection and validation}
\label{alg:dpp-injection}
\begin{algorithmic}[1]
\Require Clean model $\theta$, configuration $\mathcal{C}$, seeds $\mathcal{S}$
\Ensure Label $\mathit{isKilled}$ and aggregated run measurements $\mathbf{x}$
\State $\theta' \gets \mathrm{Inject}(\theta, \mathcal{C})$ \Comment{static or dynamic mutation}
\If{$\mathrm{VerifyStructural}(\theta, \theta', \mathcal{C}) = \mathit{false}$} \Return \textit{discard} \EndIf
\For{$s \in \mathcal{S}$}
    \State $a^c_s \gets \mathrm{TrainAndEvaluate}(\theta, s)$; \quad $a^f_s \gets \mathrm{TrainAndEvaluate}(\theta', s)$
\EndFor
\State $p \gets \mathrm{SignFlipPermTest}(\{a^c_s\}, \{a^f_s\})$ \Comment{one-sided, paired}
\State $\mathit{killed} \gets [\,p \le \alpha\,]$ \Comment{\cref{eq:dpp-iskilled}}
\State $\mathbf{x} \gets \mathrm{Aggregate}(\{a^c_s\}, \{a^f_s\})$ \Comment{clean-to-faulty deltas}
\State \Return $(\mathit{killed},\, \mathbf{x})$
\end{algorithmic}
\end{algorithm}

We fine-tune the clean model $N$ (parameters $\theta$) and the mutant $M$ (parameters $\theta'$) each $n$ times under matched random seeds, and we compare their performance distributions $A_N(\mathit{TestS}) = \langle A_{N_1}, \ldots, A_{N_n}\rangle$ and $A_M(\mathit{TestS}) = \langle A_{M_1}, \ldots, A_{M_n}\rangle$ on a held-out test set $\mathit{TestS}$. The predicate $\mathit{isKilled}$ decides whether the mutant is killed:
\begin{equation}
\mathit{isKilled}(N, M, \mathit{TestS}) =
\begin{cases}
\mathit{true}  & \text{if } p\text{-value}\bigl(A_N(\mathit{TestS}),\, A_M(\mathit{TestS})\bigr) < \alpha, \\
\mathit{false} & \text{otherwise.}
\end{cases}
\label{eq:dpp-iskilled}
\end{equation}

We use accuracy as the test-set metric for encoder classification tasks and log-perplexity for decoder language-modeling tasks~\citep{ABNN_sigma}, and we evaluate each clean and faulty configuration with the same $n=5$ seeds at $\alpha=0.05$. The matched seeds let us compare each faulty run directly with its clean counterpart. For each mutation, we compute the five paired differences and test whether they move consistently in the expected direction with a one-sided paired sign-flip permutation test~\citep{good2005permutation}. This test asks whether the observed direction of the paired differences is unlikely under the null hypothesis that the fault has no effect. With five paired comparisons, the smallest possible one-sided exact $p$-value is $1/2^5 \approx 0.031$, which lies below $\alpha=0.05$, so five matched seeds suffice for this exact test. To confirm that the test does not flag ordinary training noise as a fault, we also apply $\mathit{isKilled}$ to pairs of clean runs that differ only in random seed, and the rate at which these clean-versus-clean pairs are flagged estimates the false-positive rate. We summarize each operator's effectiveness with the mutation score~\citep{deepcrime}:

\begin{equation}
\mathit{MS}(\mathit{MO}) = \frac{|\{c \in \mathit{MO} \,:\, \mathit{isKilled}(N, M_c, \mathit{TestS})\}|}{|\mathit{MO}|}
\label{eq:dpp-ms}
\end{equation}

\noindent where $\mathit{MO}$ is the set of injected configurations of an operator, and we report the overall mutation score as the average of $\mathit{MS}(\mathit{MO})$ across operators. For each killed mutant we record a label $y = (u, f, v, \ell, \sigma)$ with the target unit, fault category, fault variant, affected layer set, and severity. We omit $m$ and $t$ because each instance comes from one model--task context. A surviving mutant, for which $\mathit{isKilled}$ returns false, is an injected fault whose measured effect does not pass the killing test, so its label cannot be inferred reliably. We discard surviving mutants before building the final benchmark instances.

\begin{figure}[htbp]
    \centering
    \includegraphics[width=0.90\linewidth]{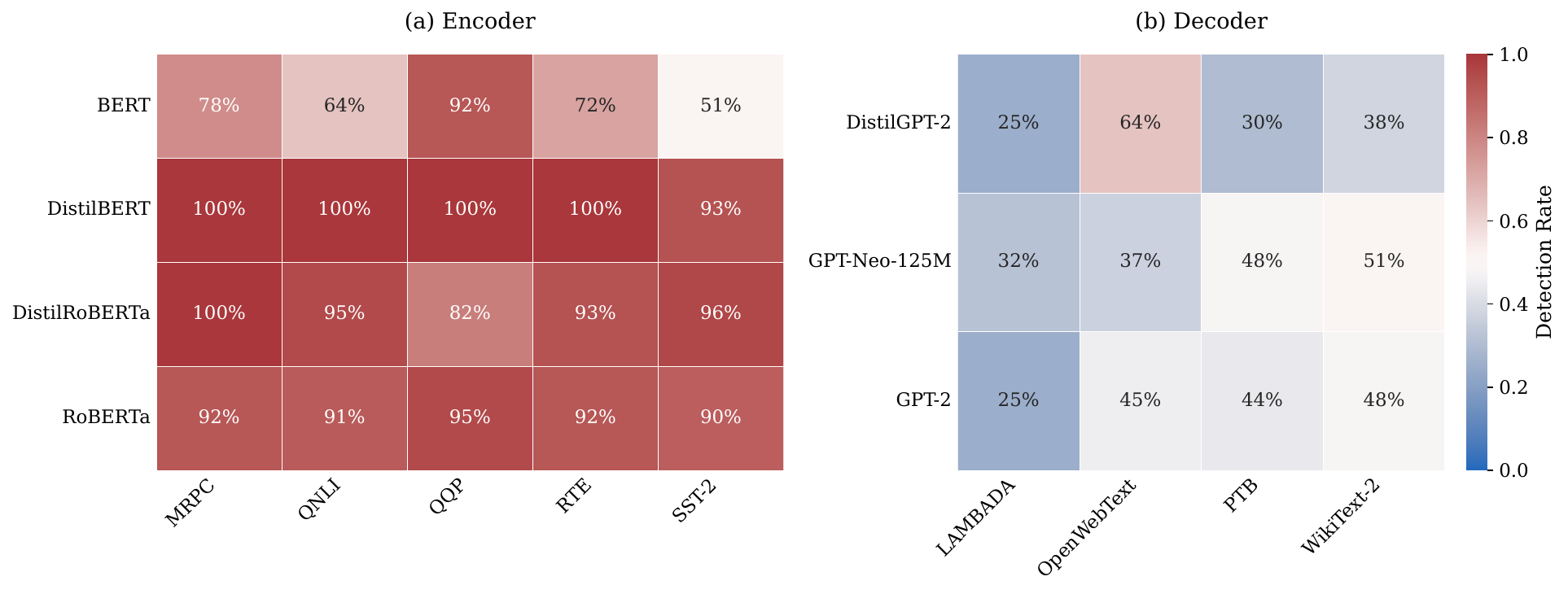}
    \caption[Mutation scores per (model, task) pair]{Mutation scores per (model, task) pair under $\mathit{isKilled}$ at $\alpha = 0.05$}
    \Description{Heatmap showing mutation scores per model and task pair. Each cell contains the fraction of injected configurations killed by the task-performance criterion.}
    \label{fig:dpp-model-task-heatmap}
\end{figure}

\subsection{Benchmark Statistics}
\label{sec:defaultpp_benchmark_statistics}
\begin{figure}[ht]
\centering
\includegraphics[width=\linewidth]{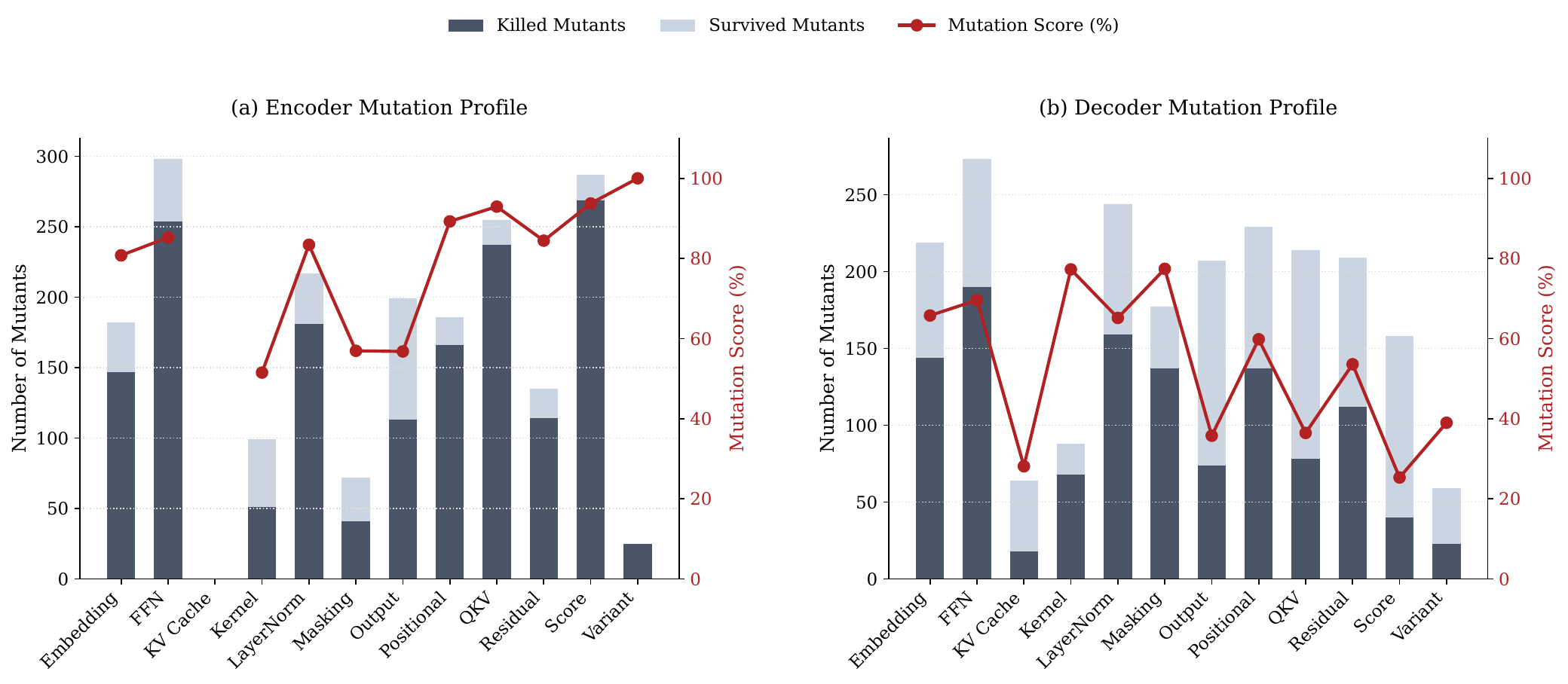}
\caption{Per-category mutation scores for encoder and decoder mutants in DEFault-bench}
\Description{Bar chart showing mutation scores for each category for encoder and decoder mutants.}
\label{fig:mutation_profiles}
\end{figure}

After structural verification, we obtain 4{,}096 single-fault configurations (1{,}955 for encoders and 2{,}141 for decoders), with every category represented. Validating these mutants requires 40{,}960 paired clean and faulty fine-tuning runs, and constructing the balanced correct class adds 13{,}890 clean-variant runs, for a total of 54{,}850 fine-tuning runs and about 27{,}000 GPU-hours on NVIDIA A100 and H100 GPUs. We apply $\mathit{isKilled}$ to all structurally valid configurations and compute the mutation score by category and architecture (\cref{tab:benchmark_summary}, \cref{fig:mutation_profiles}). Among encoder configurations, 1{,}598 of 1{,}955 mutants are killed (81.7\%). Among decoder configurations, 1{,}180 of 2{,}141 mutants are killed (55.1\%). Encoder mutation scores are high across most categories. Variant reaches 100\%, and Score, QKV, Positional, FFN, Residual, LayerNorm, and Embedding all exceed 80\%, while Kernel, Output, and Masking are the lowest. Decoder mutation scores vary more across categories and model--task pairs (\cref{fig:dpp-model-task-heatmap}). Masking and Kernel reach the highest decoder scores, and Score the lowest. The low Score rate is consistent with how softmax dampens score-level changes. Softmax maps the pre-softmax logits to a normalized distribution, so a moderate score change can be partly absorbed before it reaches the attention weights~\citep{vaswani2017attention,dong2021attention,zhai2023stabilizing}. For decoder language modeling, this can keep the perplexity change below the threshold the killing test uses.

\begin{table}[htbp]
\centering
\caption{Composition of DEFault-bench. Each value is the number of killed mutants in a fault category. Encoders contribute 1{,}598 faulty and 1{,}598 correct instances (3{,}196 total), and decoders 1{,}180 faulty and 1{,}180 correct (2{,}360 total).}
\label{tab:benchmark_summary}

\small
\setlength{\tabcolsep}{6pt}
\renewcommand{\arraystretch}{0.95}

\begin{tabular}{@{}lrrrrrrrrrrrr@{}}
\toprule
Category
& Variant
& Score
& QKV
& Pos.
& FFN
& Residual
& LayerNorm.
& Emb.
& Out.
& Mask
& Kernel
& KV \\
\midrule
Encoder
& 25
& 269
& 237
& 166
& 254
& 114
& 181
& 147
& 113
& 41
& 51
& n/a \\
Decoder
& 23
& 40
& 78
& 137
& 190
& 112
& 159
& 144
& 74
& 137
& 68
& 18 \\
\bottomrule
\end{tabular}
\end{table}

After discarding surviving mutants, we build the \textit{correct} class from clean runs and label-preserving clean variants. Using a single clean run per base model could let a classifier learn model-specific artifacts rather than fault-related behavior. We therefore generate clean variants by changing factors that should not affect task behavior, such as the random seed, the training-data order where applicable, and hyperparameters within ranges that stay below the significance threshold. We evaluate each clean variant against its base model with the same significance test used for killed mutants, retain variants that remain statistically indistinguishable from the base model, and discard variants that meet the killed criterion. For each base model $B$ that produces $k$ killed mutants, we keep $k$ clean variants from the same base model, which holds the faulty-to-correct ratio balanced within each base-model stratum. DEFault-bench contains 5{,}556 instances, with 3{,}196 encoder and 2{,}360 decoder instances (\cref{tab:benchmark_summary}).

Each instance carries labels at three levels. The Level~1 label for fault detection is binary, either \textit{faulty} or \textit{correct}. Faulty instances also carry Level~2 and Level~3 labels. The Level~2 label is the injected fault category. Encoders use 11 categories, namely Score, Positional, FFN, QKV, LayerNorm, Residual, Masking, Embedding, Kernel, Output, and Variant, and decoders use the same 11 plus KV Cache. The Level~3 label is the injected root cause within the fault category. Each category contains 2--5 root causes, and each root cause belongs to exactly one category.

\section{DEFault++~:~Hierarchical Diagnostic Technique}
\label{sec:defaultpp_approach}

\begin{figure}[htbp]
\centering
\includegraphics[width=0.85\linewidth]{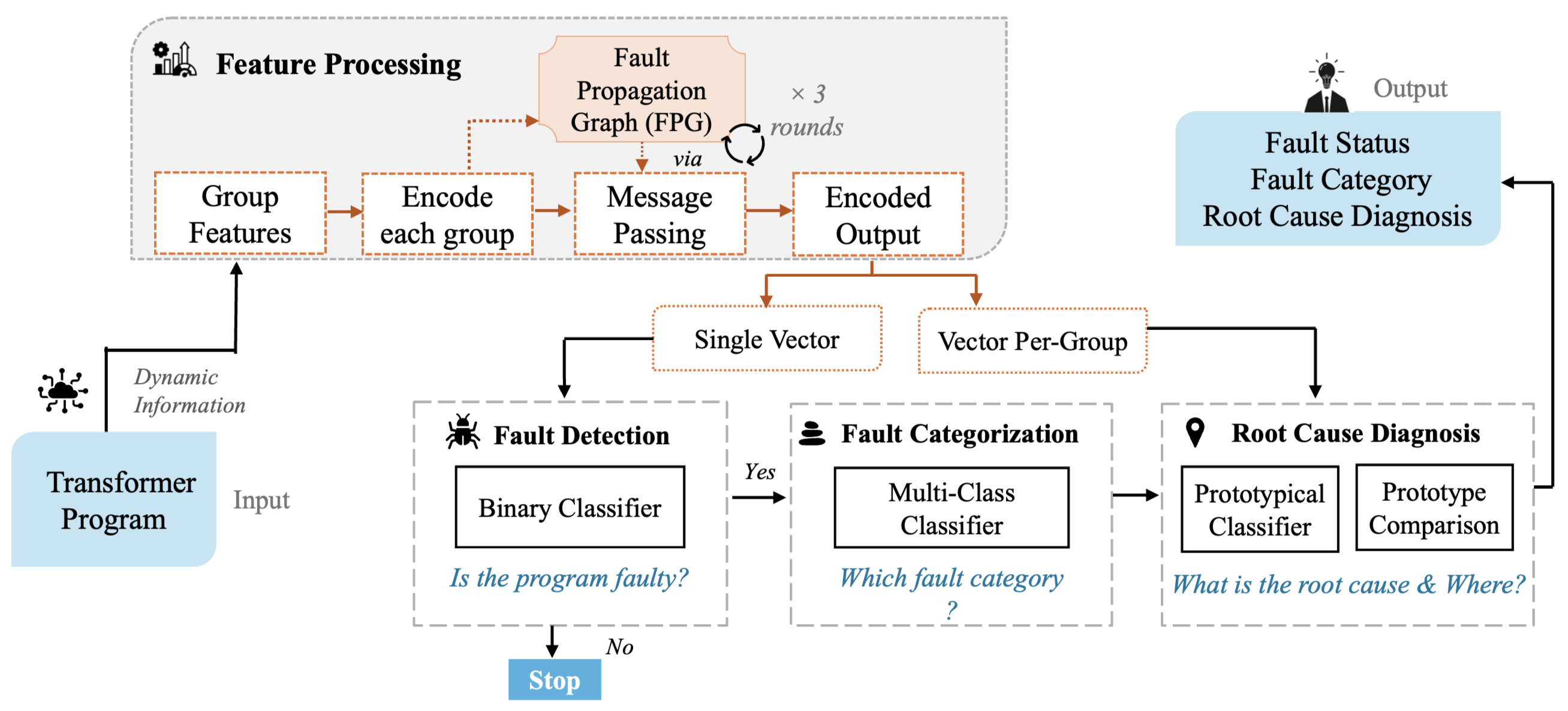}
\caption{Overview of DEFault++ for fault detection, categorization, and root-cause diagnosis}
\Description{Diagram showing the DEFault++ for fault detection, categorization, and root-cause diagnosis}
\label{fig:defaultpp_technique}
\end{figure}

DEFault++ takes the 5{,}556 labeled instances from DEFault-bench as input, each with the runtime measurements collected during training, and diagnoses them through the three-level hierarchy in \cref{fig:defaultpp_technique}. Level~1 uses the full feature representation to detect whether an instance is faulty. Only instances predicted as faulty pass to Level~2, which assigns a fault category. Level~3 then predicts the root cause within that category by comparing the instance to learned class prototypes, in an embedding space organized by the Fault Propagation Graph (FPG), the structural prior over transformer components defined in \cref{sec:defaultpp_fpg}. The same embedding space produces the feature-group importance scores that explain each diagnosis.

\subsection{Fault Propagation Graph (FPG)}
\label{sec:defaultpp_fpg}
\begin{figure}[htbp]
\centering
\includegraphics[width=0.80\linewidth]{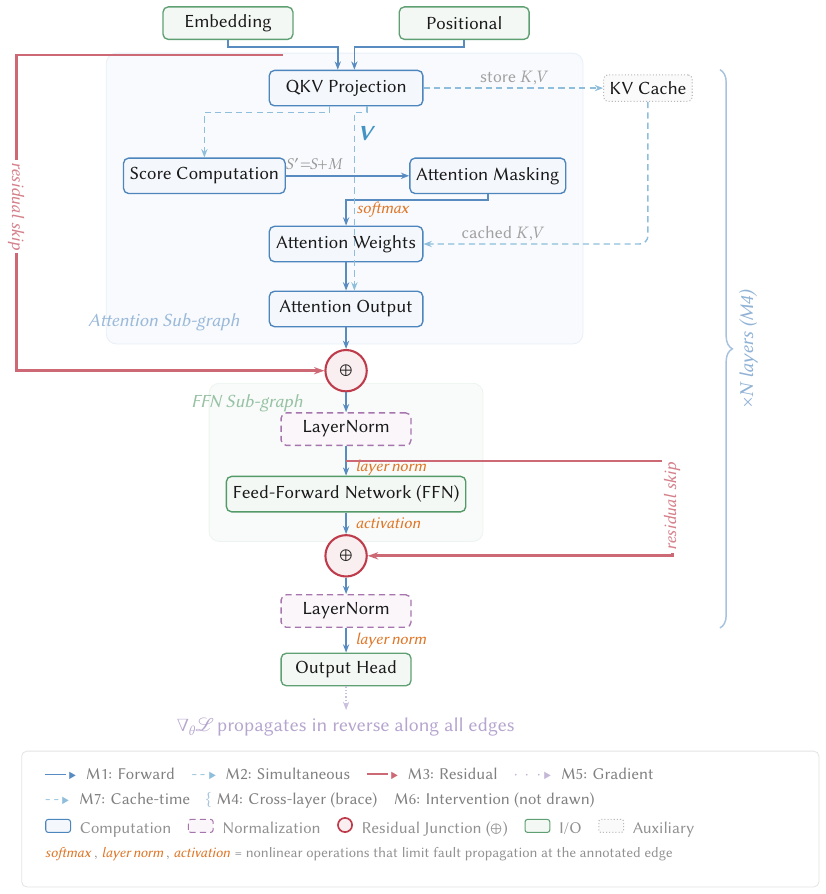}
\caption{Fault Propagation Graph (FPG) for transformer architectures}
\Description{A diagram showing the Fault Propagation Graph for transformer architectures, with labeled components and edges indicating propagation paths.}
\label{fig:defaultpp_fpg}
\end{figure}

The fault categories and root causes in our benchmark identify \emph{where} a fault originates. They do not, however, explain how the effects of a fault can propagate to other transformer components. For example, a QKV projection fault can change the projection output and then affect attention scores, attention weights, and the residual stream. To represent these dependency paths, we define the \emph{Fault Propagation Graph (FPG)}, a directed graph whose nodes are transformer components and whose edges represent data-flow dependencies (\cref{fig:defaultpp_fpg}). The FPG is deterministic because it follows the transformer computation graph. Each edge marks a possible path through which a fault can affect another component. The magnitude and representation of that effect can still change as it passes through learned transformations and operations such as softmax, LayerNorm, and activation functions.

We define the FPG as $\mathcal{G} = (\mathcal{V}, \mathcal{E})$. Each node $v \in \mathcal{V}$ represents a transformer component. Residual connections and LayerNorm appear separately for the attention and FFN sublayers. Decoder architectures also include a KV cache node. Each directed edge $(v_i, v_j) \in \mathcal{E}$ marks a dependency derived from the forward equations, residual structure, or autoregressive state reuse.

We derive the edges by analyzing the transformer's forward and backward equations, which yield seven dependency mechanisms (\cref{fig:defaultpp_fpg}). We write $\delta x$ for a small change to a variable $x$, and each mechanism below describes how such a change moves from one component to another.
\begin{enumerate}[nosep]
\item \textbf{M1: Forward sequential propagation.} If $B = f(A)$, then a change in $A$ can change $B$ according to $\delta B = (\partial f / \partial A)\, \delta A$. This mechanism captures the main forward path through embeddings, attention computation, residual connections, LayerNorm, FFN, and the output head.

\item \textbf{M2: Simultaneous propagation.} QKV projections feed both score computation (via $Q, K$) and attention output (via $V$). A single projection fault can therefore affect scores and values at the same time. In decoders, $K$ and $V$ also enter the cache.

\item \textbf{M3: Residual bypass.} In $h = x + \text{MHA}(x)$, the skip path carries changes in $x$ directly to the residual output. Later normalization steps can still adjust the magnitude or representation of the propagated change.

\item \textbf{M4: Cross-layer propagation.} The residual stream provides repeated identity paths across stacked layers~\citep{roquet2024cross}. A change at layer $\ell$ can therefore reach later representations, although intermediate normalization and nonlinear operations may reduce or reshape it.

\item \textbf{M5: Backward pass propagation.} A fault that changes the loss $\mathcal{L}$ can change $\partial \mathcal{L} / \partial \theta_i$ for parameters whose computation contributes to that loss, which can affect weight updates across multiple components during training.

\item \textbf{M6: Architecture-wide intervention.} Some fault types affect multiple components at once. A variant fault (\eg single-head instead of multi-head attention) changes the attention computation pattern, parameter shapes, and kernel dispatch. This mechanism is a joint structural effect rather than a propagation relation.

\item \textbf{M7: Cache propagation across time steps.} In decoder architectures, the KV cache stores $K$ and $V$ across generation steps. A cache fault can affect the current token and future tokens.
\end{enumerate}

Three nonlinear operations limit propagation along specific edges. Softmax maps changes into a normalized distribution and reduces large shifts. LayerNorm~\citep{ba2016layer} rescales activations by their mean and variance. Activation functions such as ReLU and GELU suppress or pass changes depending on their operating region~\citep{wang2019reltanh, He2020Piecewise}. We annotate these limiting operations on the corresponding FPG edges. The seven mechanisms fall into three classes by how the diagnostic model uses each one, namely \emph{propagation relations} that become graph edges, \emph{recursive composition} that repeats across stacked layers, and a \emph{structural intervention} that the model reads from the fault labels rather than from edges (\cref{tab:fpg_derivation}). The propagation relations (M1, M2, M3, M5, M7) describe paths through which a change reaches another component, and the message-passing graph (\cref{sec:defaultpp_reasoning}) uses the forward relations (M1, M2, M3, M7) as edges while representing the backward relation (M5) through gradient features. Recursive composition (M4) repeats M1 and M3 across stacked layers, and the structural intervention (M6) is a joint effect in which one architectural change alters several components at once.

\begin{table}[htbp]
\caption{Edge derivation for the Fault Propagation Graph (FPG)}
\label{tab:fpg_derivation}
\centering
\scriptsize
\renewcommand{\arraystretch}{1.0}
\setlength{\tabcolsep}{4pt}
\begin{tabularx}{\textwidth}{@{} p{2.2cm} p{2.3cm} p{2.8cm} p{1.3cm} X @{}}
\toprule
\textbf{Source} & \textbf{Target} & \textbf{Mechanism Class} & \textbf{Scope} & \textbf{Details} \\
\midrule
Embedding & QKV Projection & Forward prop.\ (M1) & Enc/Dec & $Q,K,V = W_{Q,K,V}\, E[x]$ \\
Positional & QKV Projection & Forward prop.\ (M1) & Enc/Dec & Position added to embedding before projection \\
QKV Proj.\ & Score Comp. & Simultaneous (M2) & Enc/Dec & $S = QK^\top\!/\sqrt{d_k}$; $Q,K$ from projection \\
QKV Proj.\ & Attn.\ Output & Simultaneous (M2) & Enc/Dec & $\text{Attn} = \text{softmax}(S)\,V$; $V$ from projection \\
QKV Proj.\ & KV Cache & Simultaneous (M2) & Dec & $K,V$ stored in cache at each step \\
Score Comp.\ & Attn.\ Mask & Forward prop.\ (M1) & Enc/Dec & $S' = S + M$ \\
Attn.\ Mask & Attn.\ Weights & Forward prop.\ (M1) & Enc/Dec & $\alpha = \text{softmax}(S')$ \\
Attn.\ Weights & Attn.\ Output & Forward prop.\ (M1) & Enc/Dec & $\text{Attn} = \alpha\, V$ \\
Attn.\ Output & Residual (attn) & Forward prop.\ (M1) & Enc/Dec & $h = x + \text{Attn}(x)$ \\
Residual input & Residual (attn) & Residual bypass (M3) & Enc/Dec & Skip connection: $\delta x$ passes with unit gain \\
Residual (attn) & LayerNorm (attn) & Forward prop.\ (M1) & Enc/Dec & $\hat{x} = \text{LN}(h)$ \\
LayerNorm (attn) & FFN & Forward prop.\ (M1) & Enc/Dec & FFN input is post-norm representation \\
FFN & Residual (FFN) & Forward prop.\ (M1) & Enc/Dec & $h' = h + \text{FFN}(\hat{x})$ \\
Residual input & Residual (FFN) & Residual bypass (M3) & Enc/Dec & Skip connection: $\delta h$ passes with unit gain \\
Residual (FFN) & LayerNorm (FFN) & Forward prop.\ (M1) & Enc/Dec & $\hat{h}' = \text{LN}(h')$ \\
LayerNorm (FFN) & Output Head & Forward prop.\ (M1) & Enc/Dec & Final representation enters output head \\
Layer $\ell$ residual & Layer $\ell{+}1$ & Cross-layer (M4) & Enc/Dec & Repeated M1+M3 across stacked layers \\
n/a & n/a & Backward prop.\ (M5) & Enc/Dec & Represented through gradient features \\
KV Cache & Attn.\ Weights & Cache-time (M7) & Dec & Cached $K,V$ affect future-step attention \\
n/a & n/a & Intervention (M6) & Enc/Dec & Joint multi-component effect, not represented as a propagation edge \\
\bottomrule
\end{tabularx}
\end{table}


\subsection{Feature Representation}
\label{sec:defaultpp_features}
\begin{figure}[htbp]
\centering
\includegraphics[width=\linewidth]{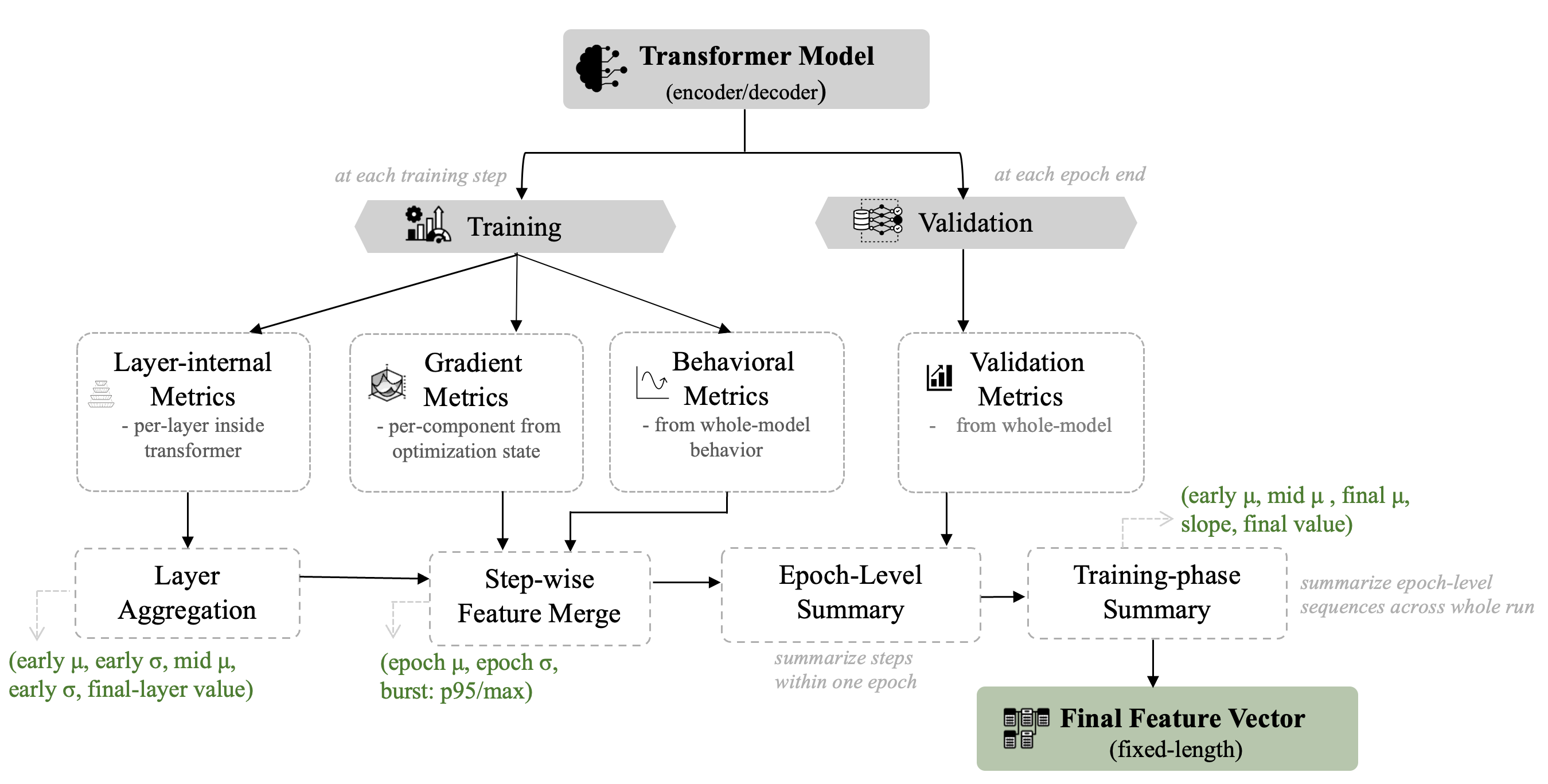}
\Description{Feature construction process from raw metrics to a fixed-length feature vector}
\caption{DEFault++ feature construction flow.}
\label{fig:feature_technique}
\end{figure}

We represent each training run with features that capture component-level behavior. The FPG identifies the component paths along which a fault can propagate, and we use this structure to organize the collected metrics. Architecture-agnostic DNN metrics, such as loss trajectories, accuracy, and gradient statistics, are often too broad to separate transformer-specific fault categories. A fault in a QKV projection, a mask, or a positional encoding may barely move accuracy or perplexity while clearly changing attention-level and component-level measurements. We therefore collect metrics for attention distributions, projection alignment, residual-stream behavior, optimization behavior, training dynamics, and validation behavior, and we organize them into \emph{feature groups}, where each feature group holds the metrics for one transformer component or one model-wide aspect of training (\cref{fig:feature_technique}, \cref{tab:defaultpp_features}).

By how they are measured, these feature groups fall into four families, and \cref{tab:defaultpp_features} lists the metric types in each group, where $n_g$ is the number of metric types in group $g$ before aggregation. \emph{Layer-internal} metrics capture attention and component behavior inside transformer layers, since attention distributions~\citep{voita2019analyzing,clark2019does}, representation geometry~\citep{ethayarajh2019contextual,kornblith2019similarity}, normalization parameters~\citep{ba2016layer,xiong2020layer}, and residual propagation~\citep{roquet2024cross} can each vary across layers. They total $C_{int}=15$ metric types for encoders and $C_{int}=16$ for decoders, where the decoder adds future attention mass. \emph{Gradient} metrics total $C_{opt}=21$ types and record gradient norm, update ratio, and update activity across parameterized components and global gradients~\citep{default,deepfd,autotrainer,deeplocalize}. \emph{Behavioral} metrics total $C_{train}=10$ types for encoders and $12$ for decoders and capture output-head behavior, positional sensitivity, step time, peak memory allocation, and decoder cache behavior. \emph{Validation} metrics total $C_{eval}=2$ types and capture task accuracy or perplexity and calibration error at validation checkpoints~\citep{guo2017calibration}.

\begin{table}[htbp]
\caption{Feature groups, metric types, and pre-aggregation feature counts $n_g$ in DEFault++. Here $n_g$ is the number of scalar metric types in a group before aggregation. The attention group has six metric types for encoders and seven for decoders, since future attention mass applies only to decoders, and gradient metrics are routed to their component groups.}
\label{tab:defaultpp_features}
\centering
\small
\renewcommand{\arraystretch}{1.00}
\setlength{\tabcolsep}{4pt}

\begin{tabularx}{\linewidth}{@{} p{2.3cm} X c @{}}
\toprule
\textbf{Feature Group} & \textbf{Metric Types} & $\mathbf{n_g}$ \\
\midrule

\multicolumn{3}{c}{\textit{Layer-internal metrics ($C_{int}$, collected at each training step, per layer)}} \\[2pt]

Attention & Attention entropy, padding attention mass, inter-head cosine similarity, head usage, attention rank, cross-example leakage, future attention mass (decoder only) & 6/7 \\[2pt]

Score & Pre-softmax score & 1 \\[2pt]

FFN output & FFN output norm & 1 \\[2pt]

LayerNorm & Scale parameter norm, post-norm distribution & 2 \\[2pt]

Residual stream & Residual cosine similarity & 1 \\[2pt]

Repr.\ drift & CKA layer similarity & 1 \\[2pt]

QKV alignment & Q--K sim., Q--V sim., K--V sim. & 3 \\[2pt]

\midrule
\multicolumn{3}{c}{\textit{Gradient metrics ($C_{opt}$, collected at each training step)}} \\[2pt]

Optimization & Gradient and update features (gradient norm, update ratio, update activity) & 21 \\[2pt]

\midrule
\multicolumn{3}{c}{\textit{Behavioral metrics ($C_{train}$, collected at each training step)}} \\[2pt]

Embedding & Embedding norm, token-level variance & 2 \\[2pt]

Positional & Positional sensitivity & 1 \\[2pt]

Training dyn. & Loss trajectory, gradient norm volatility, step time, peak memory allocation & 4 \\[2pt]

Output & Prediction confidence, output entropy, margin statistics & 3 \\[2pt]

Cache & Cache hidden similarity, cache distribution divergence & 2 \\[2pt]

\midrule
\multicolumn{3}{c}{\textit{Validation metrics ($C_{eval}$, collected at epoch end or validation checkpoints)}} \\[2pt]

Validation perf. & Task accuracy/perplexity, calibration error & 2 \\

\bottomrule
\end{tabularx}
\end{table}

Training runs can differ in model depth, number of steps, and number of epochs. We therefore aggregate the raw measurements at the layer, epoch, and training-phase levels, as shown in \cref{fig:feature_technique}. Layer aggregation reduces each per-layer metric to $A_\ell=5$ statistics, preserving early, middle, and final-layer behavior while removing dependence on model depth. Epoch-level aggregation reduces the step-level sequence within each epoch to $A_e=3$ statistics. Training-phase aggregation reduces the epoch sequence to $A_p=5$ statistics. This aggregation produces one feature vector per training run while retaining information about the affected component and the training phase in which the effect appears. Combining the per-family metric-type counts with these aggregation factors gives the length of the final feature vector,

\begin{equation}
d_{\text{final}} = A_p\!\left(A_e\!\left(A_\ell \, C_{int} + C_{opt} + C_{train}\right) + C_{eval}\right),
\label{eq:feature_vector_length}
\end{equation}

\Cref{eq:feature_vector_length} gives $1{,}600$ features for encoders ($C_{int} = 15$, $C_{opt} = 21$, $C_{train} = 10$, $C_{eval} = 2$) and $1{,}705$ features for decoders, where $C_{int} = 16$ adds future attention mass and $C_{train} = 12$ adds two cache metrics. We remove near-constant columns using coefficient-of-variation filtering ($\mathrm{CV} < 0.01$). DEFault++ uses the filtered fixed-length feature vector as input to the diagnostic model.

\subsection{Diagnostic Model}
\label{sec:defaultpp_reasoning}

DEFault++ performs hierarchical diagnosis for one input instance at a time. Level~1 detects whether the instance is faulty. If Level~1 predicts a fault, Level~2 assigns the instance to a fault category. Level~3 then predicts the root cause within that category and computes feature-group importance scores for the diagnosis. \\

\textbf{Feature-Group Encoding.} DEFault++ first encodes each feature group (\cref{tab:defaultpp_features}) with a dedicated multilayer perceptron (MLP), which keeps the component-level organization of the measurements. Encoders and decoders have different numbers of feature groups, $G = 12$ for encoders and $G = 13$ for decoders, since the KV cache group applies only to decoders, so we train a separate diagnostic model for each architecture.

Given a feature vector $\mathbf{z} \in \mathbb{R}^d$ partitioned into $G$ groups $\{\mathbf{z}_{g_1}, \ldots, \mathbf{z}_{g_G}\}$ (see \cref{tab:defaultpp_features}), each group $g$ is encoded as
\begin{equation}
\mathbf{h}_g = \text{MLP}_g(\mathbf{z}_g) \in \mathbb{R}^{h}, \quad g = 1, \ldots, G,
\label{eq:group_encoding}
\end{equation}
where $h = 32$ is the hidden dimension per group, selected by grid search on the inner validation fold. The group embeddings form the matrix $\mathbf{H} = [\mathbf{h}_1; \ldots; \mathbf{h}_G] \in \mathbb{R}^{G \times h}$.

A learnable projection matrix $W_{\text{proj}}$ then maps the concatenated group embeddings to a single shared representation,
\begin{equation}
\mathbf{z}_{\text{proj}} = W_{\text{proj}} \, \text{vec}(\mathbf{H}) \in \mathbb{R}^{e},
\label{eq:projection}
\end{equation}
where $e = 64$ is the embedding dimension and $\text{vec}(\mathbf{H}) \in \mathbb{R}^{Gh}$ denotes the row-major vectorization of $\mathbf{H}$. Levels~1 and~2 use this shared vector $\mathbf{z}_{\text{proj}}$ for fault detection and fault categorization, while Level~3 uses the group embedding matrix $\mathbf{H}$ to preserve group-level structure for root-cause diagnosis and explanation.\\

\textbf{Message Passing in the Fault Propagation Graph (FPG).}
We use the FPG to exchange information between related feature groups. Each round updates a feature group's embedding using its current value and the embeddings of the feature groups connected to it by FPG edges. After several rounds, a group embedding includes evidence from nearby groups in the FPG. For example, the QKV alignment group can aggregate evidence from the score, attention, and residual-stream groups, which lie on the propagation paths of QKV projection faults.

The group-level graph uses the forward propagation mechanisms defined in the FPG. Backward effects are captured by gradient-based feature groups instead of graph edges. We construct the adjacency matrix $\mathbf{A} \in \mathbb{R}^{G \times G}$ by mapping FPG components to their corresponding feature groups (\cref{tab:fpg_mapping}). A group-level edge exists when any component in the source group has an FPG edge to any component in the target group. Components mapped to the same feature group are merged, and groups that cover multiple components inherit the edges of those components. Representation drift, Training dynamics, and Validation performance do not correspond to individual transformer components, so they use self-loops only.

\begin{table}[htbp]
\caption{Feature-group mapping for FPG message passing in DEFault++}
\label{tab:fpg_mapping}
\centering
\footnotesize
\renewcommand{\arraystretch}{1.0}
\setlength{\tabcolsep}{7pt}
\begin{tabular}{@{} l l l l @{}}
\toprule
\textbf{Feature Group} & \textbf{Role} & \textbf{Scope} & \textbf{Message Passing} \\
\midrule
Attention & Structural & Attention masking, weights, output & FPG edges \\
Score & Structural & Score computation & FPG edges \\
FFN output & Structural & FFN & FPG edges \\
LayerNorm & Structural & LayerNorm & FPG edges \\
Residual stream & Structural & Residual connections & FPG edges \\
QKV alignment & Structural & QKV projection & FPG edges \\
Embedding & Structural & Embedding & FPG edges \\
Positional & Structural & Positional encoding & FPG edges \\
Output & Structural & Output head & FPG edges \\
Cache (decoder) & Structural & KV cache & FPG edges \\
\midrule
Representation drift & Cross-layer observation & Inter-layer residual path & Self-loop only \\
\midrule
Training dynamics & Model-wide context & Model-wide optimization & Self-loop only \\
Validation performance & Model-wide context & Model-wide output quality & Self-loop only \\
\bottomrule
\end{tabular}
\end{table}

\begin{figure}[htbp]
\centering
\includegraphics[width=0.60\linewidth]{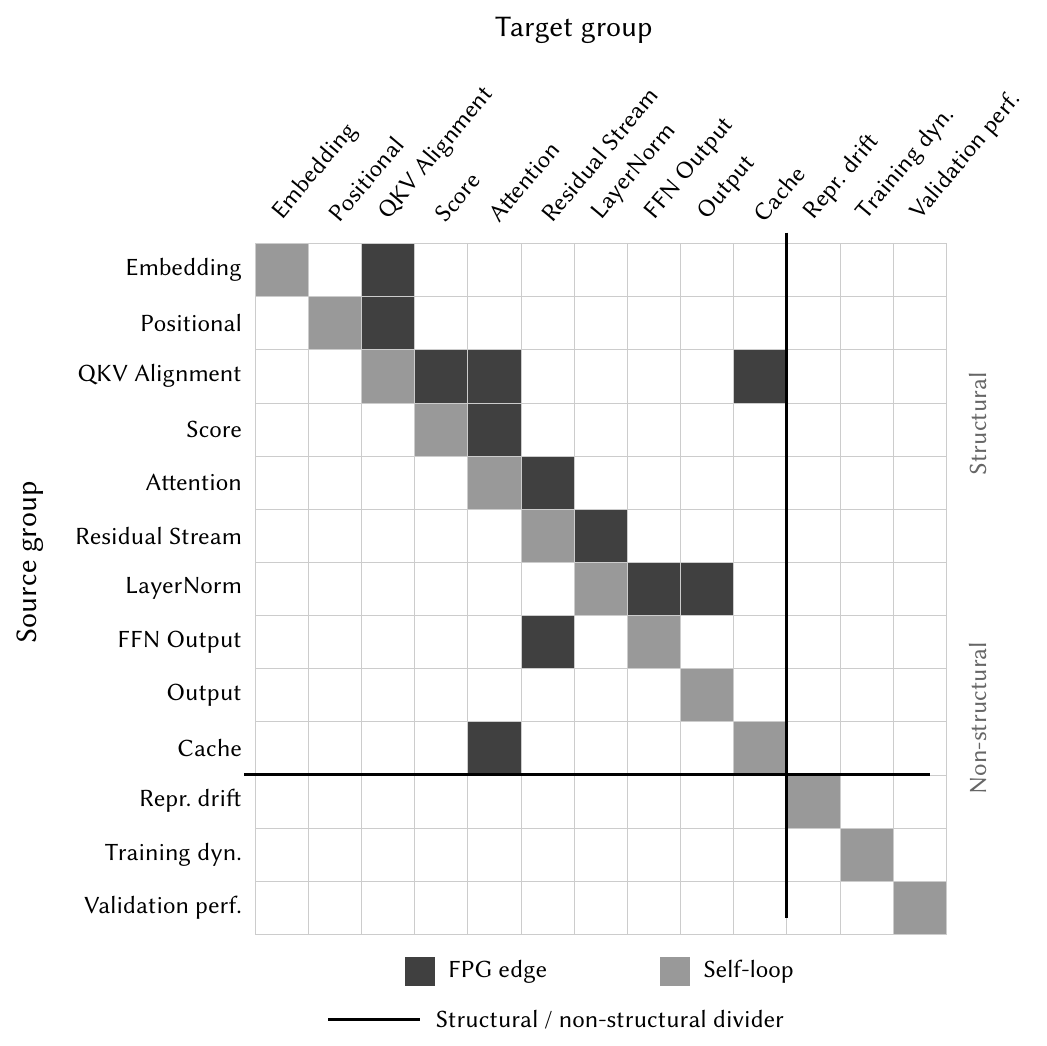}
\caption{Group-level adjacency matrix $\hat{\mathbf{A}}$ for the decoder diagnostic model}
\Description{Visualization of the group-level adjacency matrix for the decoder diagnostic model.}
\label{fig:defaultpp_adjacency}
\end{figure}

For the decoder diagnostic model ($G = 13$), \cref{fig:defaultpp_adjacency} shows the row-normalized adjacency matrix with self-loops, $\hat{\mathbf{A}}$. Nonzero values near the main diagonal follow the sequential data flow through the transformer block. Off-diagonal values capture branching paths such as QKV$\to$Cache, Cache$\to$Attention, and FFN$\to$Residual. Representation drift, Training dynamics, and Validation performance appear as isolated self-loops, since they provide model-wide context rather than lying on a component-level propagation path.

Each round mixes every group embedding with those of its FPG neighbors and then applies a learned transformation. Following graph convolutional networks~\citep{kipf2017semi}, one round of message passing is
\begin{equation}
\mathbf{H}' = \operatorname{ReLU}\left(\hat{\mathbf{A}}\mathbf{H}W_{\text{msg}}\right),
\label{eq:message_passing}
\end{equation}
\noindent where $\hat{\mathbf{A}}$ is the row-normalized adjacency matrix with self-loops and $W_{\text{msg}} \in \mathbb{R}^{h \times h}$ is a learnable weight matrix. We use row normalization because feature groups have different numbers of FPG neighbors, and it keeps the message-passing updates on a comparable scale across groups.

We apply three rounds of message passing, with a separate $W_{\text{msg}}$ for each round. We select the number of rounds by grid search over $\{1,\dots,5\}$ on the inner validation fold. Three rounds cover short FPG paths, such as QKV $\to$ Score $\to$ Attention $\to$ Residual, while avoiding deeper aggregation that can oversmooth node representations~\citep{li2018deeper}. DEFault++ uses the updated group embeddings in the following diagnosis stages.

\textbf{Hierarchical Classification.} DEFault++ uses a shared encoder across the three diagnosis levels. The encoder combines feature-group encoding, FPG message passing, and projection into a common representation. Each diagnosis level then applies its own prediction head to this representation. Sharing the encoder avoids separate representations for closely related tasks and keeps the model small, since message passing operates over only 12 feature groups for encoders and 13 for decoders.\\
\indent
\sbullet~\textit{Level~1: Fault detection.}
The first level performs binary detection. A two-layer MLP maps $\mathbf{z}_{\text{proj}}$ to logits for faulty and correct instances.\\
\indent
\sbullet~\textit{Level~2: Fault categorization.}
Instances predicted as faulty are passed to the second level, which assigns them to a fault category. A two-layer MLP maps $\mathbf{z}_{\text{proj}}$ to $C$ category logits, where $C = 11$ for encoders and $C = 12$ for decoders since the KV Cache category applies only to decoders.\\
\indent
\sbullet~\textit{Level~3: Root-cause diagnosis.}
The third level predicts the root cause within the fault category selected by Level~2. Since each category contains only 2--5 root causes and per-root-cause support is limited, DEFault++ uses a \emph{prototypical classifier}~\citep{snell2017prototypical,kamp1995prototype}, which represents each root cause by a single point in the embedding space, its \emph{prototype}, and classifies an instance by its distance to those points. A prototype is the mean group embedding of the training examples for a root cause, and at inference DEFault++ assigns the instance to the nearest root-cause prototype within the predicted category.

Let $\mathcal{D}_{c,r}$ denote the training samples with fault category $c$ and root cause $r$. For each root cause $r$ within category $c$, we compute the prototype $\symbf{\pi}_{c,r}$ as the mean group embedding:
\begin{equation}
\symbf{\pi}_{c,r} = \frac{1}{|\mathcal{D}_{c,r}|} \sum_{i \in \mathcal{D}_{c,r}} \mathbf{H}_i \in \mathbb{R}^{G \times h}.
\label{eq:prototype}
\end{equation}

DEFault++ compares an input instance with the prototypes from the predicted category. The squared distance to a prototype decomposes over feature groups:
\begin{equation}
d(\mathbf{H}, \symbf{\pi}_{c,r}) = \sum_{g=1}^{G} \| \mathbf{h}_g - \symbf{\pi}_{c,r,g} \|^2 = \sum_{g=1}^{G} d_g,
\label{eq:proto_distance}
\end{equation}
where $d_g = \| \mathbf{h}_g - \symbf{\pi}_{c,r,g} \|^2$ is the contribution of feature group $g$. DEFault++ predicts the nearest root-cause prototype:
\[
\hat{r} = \arg\min_{r \in \mathcal{R}_{\hat{c}}} d(\mathbf{H}, \symbf{\pi}_{\hat{c},r}).
\]
The group-wise decomposition of the prototype distance allows DEFault++ to estimate each feature group's contribution to the root-cause decision.

\newcommand{\LevelHeader}[1]{%
  \Statex \(\triangleright\) \textit{#1}%
}

\begin{algorithm}[!htbp]
\caption{DEFault++ hierarchical diagnosis for one input instance}
\label{alg:defaultpp_inference}
\begin{algorithmic}[1]
\Statex \textbf{Input:} Feature vector $\mathbf{x}$; shared encoder (group embeddings $\{\mathbf{h}_g\}_{g=1}^{G}$ and projection $\mathbf{z}_{\text{proj}}$); trained prototypes $\{\symbf{\pi}_{c,r}\}$
\Statex \textbf{Output:} Detection label $\hat{y}$; if faulty: fault category $\hat{c}$, root cause $\hat{r}$, and per-group importance scores $\{w_g\}_{g=1}^{G}$
\LevelHeader{Level 1: Fault Detection}
\State Classify the instance as faulty or correct: $\hat{y} \gets \text{Detect}(\mathbf{z}_{\text{proj}})$
\If{$\hat{y} = \textit{correct}$}
    \State \Return $\hat{y}$ \Comment{correct instances exit here;}
\EndIf
\LevelHeader{Level 2: Fault Categorization}
\State Assign the fault to a transformer component category: $\hat{c} \gets \text{Categorize}(\mathbf{z}_{\text{proj}})$
\LevelHeader{Level 3: Root-Cause Diagnosis}
\For{each root cause $r$ in the predicted category $\mathcal{R}_{\hat{c}}$}
    \State Sum squared distances to prototype $r$ across all feature groups: $d_r \gets \sum_{g=1}^{G} \lVert \mathbf{h}_g - \symbf{\pi}_{\hat{c},r,g} \rVert^2$
\EndFor
\State Select the nearest prototype as the predicted root cause: $\hat{r} \gets \arg\min_{r \in \mathcal{R}_{\hat{c}}}\; d_r$
\LevelHeader{Level 3: Root-Cause Explanation}
\State Identify the nearest alternative root cause: $\tilde{r} \gets \arg\min_{r \in \mathcal{R}_{\hat{c}} \setminus \{\hat{r}\}}\; d_r$
\For{each feature group $g = 1, \ldots, G$}
    \State Compute how much group $g$ favors $\hat{r}$ over $\tilde{r}$: $\Delta_g \gets d_g(\symbf{\pi}_{\hat{c},\tilde{r}}) - d_g(\symbf{\pi}_{\hat{c},\hat{r}})$
    \State Normalize positive contributions into importance score: $w_g \gets \max(\Delta_g,\,0)\;/\;\sum_{g'} \max(\Delta_{g'},\,0)$
\EndFor
\State \Return $\hat{y},\; \hat{c},\; \hat{r},\; \{w_g\}_{g=1}^{G}$
\end{algorithmic}
\end{algorithm}

\subsection{Training Objective}
\label{sec:defaultpp_training}
DEFault++ trains the three diagnosis levels jointly with a shared encoder. During training, the detection loss, categorization loss, and root-cause diagnosis losses all update the shared encoder.

The training objective has four components. Level~1 uses binary cross-entropy, $\mathcal{L}_{\text{detect}}$, over all samples in the batch. Level~2 uses class-weighted cross-entropy, $\mathcal{L}_{\text{cat}}$, over fault categories and applies it only to faulty samples. Level~3 uses two losses, a per-category cross-entropy loss $\mathcal{L}_{\text{rc}}$ for root-cause prediction and a separation loss $\mathcal{L}_{\text{sep}}$ for prototype-based diagnosis.

For root-cause training, DEFault++ learns root-cause distinctions separately within each fault category. For each fault category, DEFault++ uses a separate root-cause head. The head takes $\mathbf{z}_{\text{proj}}$ as input and predicts among the root causes in that category. For example, QKV root causes are compared with other QKV root causes, not with LayerNorm or FFN root causes. At inference, the model first predicts a fault category and then diagnoses the root cause within that category. We compute $\mathcal{L}_{\text{rc}}$ only over faulty samples within their ground-truth fault category.

\begin{figure}[htbp]
\centering
\includegraphics[width=\linewidth]{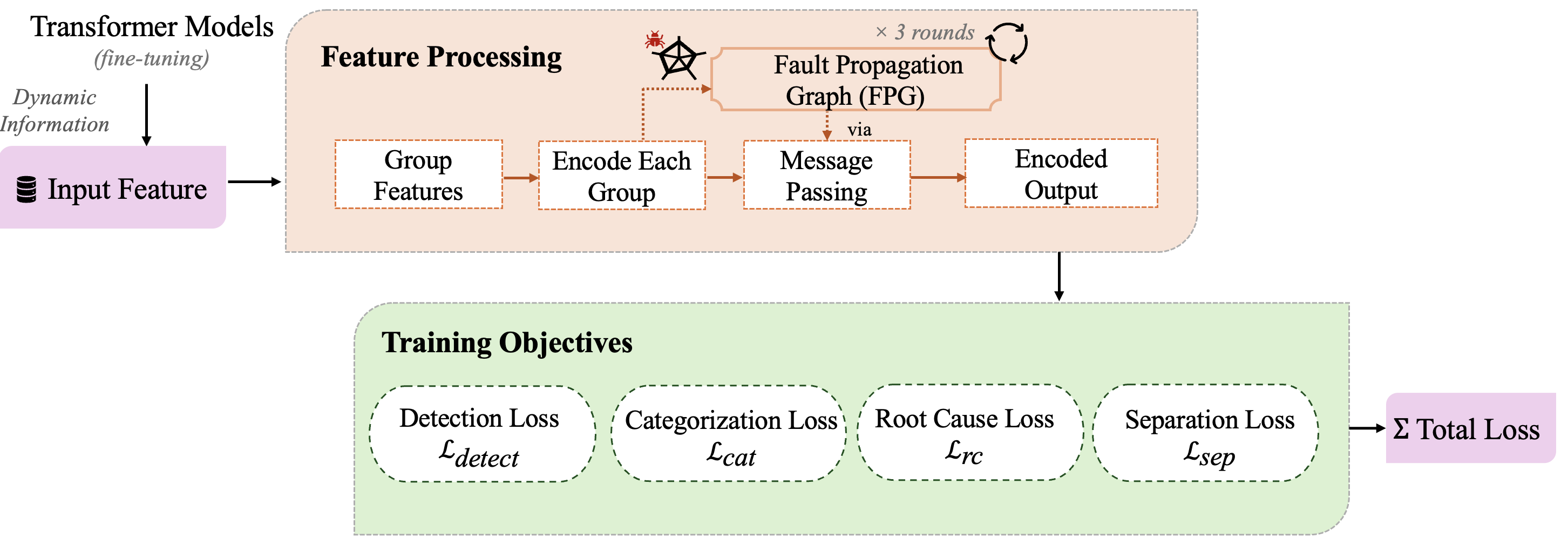}
\caption{Training view of DEFault++}
\Description{Diagram showing how DEFault++ processes features and trains its four loss terms: detection, categorization, root cause, and separation.}
\label{fig:defaultpp_training}
\end{figure}

The cross-entropy loss used for root-cause diagnosis trains the per-category heads, but prototype-based diagnosis also depends on the group embedding space $\mathbf{H}$. Root causes from the same fault category may still have similar group embeddings, which can make nearest-prototype diagnosis less reliable. DEFault++ therefore uses $\mathcal{L}_{\text{sep}}$ to separate root causes in the group embedding space. The separation loss combines supervised contrastive learning~\citep{khosla2020supervised} with prototype matching~\citep{snell2017prototypical}. The contrastive term separates embeddings of different root causes within the same category, while the prototype-matching term moves each sample closer to its own root-cause prototype.

The contrastive term uses one vector per training sample, computed by flattening the group embedding matrix: $\mathbf{u}_i = \text{vec}(\mathbf{H}_i)$. For a training sample $i$, the comparison set is restricted to samples from the same fault category. Within this set, positives are samples with the same root cause as $i$, and negatives are samples with a different root cause. The contrastive term therefore pulls together embeddings from the same root cause and separates embeddings from different root causes within the same category. Following supervised contrastive learning~\citep{khosla2020supervised}, the contrastive term for category $c$ is
\begin{equation}
\mathcal{L}_{\text{ctr}}^{(c)} = -\frac{1}{|\mathcal{P}_c|} \sum_{(i,j) \in \mathcal{P}_c} \log \frac{\exp(\text{sim}(\mathbf{u}_i, \mathbf{u}_j) / \tau)}{\sum_{\substack{k \ne i \\ c_k = c_i}} \exp(\text{sim}(\mathbf{u}_i, \mathbf{u}_k) / \tau)}
\label{eq:contrastive}
\end{equation}
\noindent where $\mathcal{P}_c$ is the set of positive pairs within category $c$, $\text{sim}(\cdot, \cdot)$ is cosine similarity, and $\tau = 0.1$ is the temperature. The denominator excludes sample $i$ itself and sums over the other samples in the batch that share the same fault category. We compute $\mathcal{L}_{\text{ctr}}$ by averaging $\mathcal{L}_{\text{ctr}}^{(c)}$ over categories with at least one positive pair in the batch.

The prototype-matching term uses the grouped squared distance from \cref{eq:proto_distance}. It encourages each sample to lie closer to its own root-cause prototype than to other prototypes in the same category:
\begin{equation}
\mathcal{L}_{\text{pm}} = -\frac{1}{|\mathcal{B}|} \sum_{i \in \mathcal{B}} \log \frac{\exp(-d(\mathbf{H}_i, \symbf{\pi}_{c_i, r_i}) / \tau_p)}{\sum_{r' \in \mathcal{R}_{c_i}} \exp(-d(\mathbf{H}_i, \symbf{\pi}_{c_i, r'}) / \tau_p)}
\label{eq:proto_matching}
\end{equation}
\noindent where $\mathcal{B}$ is the training batch, $\mathcal{R}_{c_i}$ is the set of root causes within category $c_i$, and $\tau_p$ is a temperature parameter.

The separation loss combines these two terms:
\begin{equation}
\mathcal{L}_{\text{sep}} = \beta\,\mathcal{L}_{\text{ctr}} + \gamma\,\mathcal{L}_{\text{pm}}.
\label{eq:separation_loss}
\end{equation}

Finally, the full DEFault++ objective is
\begin{equation}
\mathcal{L}_{\text{DEFault++}} = \mathcal{L}_{\text{detect}} + \alpha\,\mathcal{L}_{\text{cat}} + \lambda\,\mathcal{L}_{\text{rc}} + \mathcal{L}_{\text{sep}}.
\label{eq:total_loss}
\end{equation}
\noindent We apply $\mathcal{L}_{\text{sep}}$ only to faulty samples within their ground-truth category. We set $\alpha = 1.0$, $\lambda = 1.0$, $\beta = 0.5$, and $\gamma = 0.3$ through grid search.

\subsection{FPG-based Explanation}
\label{sec:defaultpp_explainability}

Level~3 predicts the root cause by selecting the nearest prototype in the grouped embedding space. To explain this prediction, DEFault++ compares the predicted prototype $\symbf{\pi}_{\hat{c},\hat{r}}$ with the nearest alternative prototype $\symbf{\pi}_{\hat{c},\tilde{r}}$ in the predicted fault category:
\[
\tilde{r} = \arg\min_{r \in \mathcal{R}_{\hat{c}} \setminus \{\hat{r}\}} d(\mathbf{H}, \symbf{\pi}_{\hat{c},r}).
\]
For each feature group $g$, DEFault++ computes how much closer the group embedding is to the predicted prototype than to the nearest competing prototype:
\begin{equation}
\Delta_g = d_g(\symbf{\pi}_{\hat{c},\tilde{r}}) - d_g(\symbf{\pi}_{\hat{c},\hat{r}}), \quad g = 1, \ldots, G,
\label{eq:group_contribution}
\end{equation}
where $d_g(\symbf{\pi}_{\hat{c},r}) = \| \mathbf{h}_g - \symbf{\pi}_{\hat{c},r,g} \|^2$. A positive $\Delta_g$ means that feature group $g$ supports the predicted root cause over the nearest competing root cause. A negative $\Delta_g$ means that the group is closer to the competing prototype. Since the full prototype distance is the sum of group-level distances, $\sum_g \Delta_g$ equals the distance gap between the predicted prototype and the nearest competing prototype.

DEFault++ normalizes the positive group contributions into importance scores:
\begin{equation}
w_g = \frac{\max(\Delta_g, 0)}{\sum_{g'=1}^{G} \max(\Delta_{g'}, 0)}, \quad g = 1, \ldots, G.
\label{eq:explanation}
\end{equation}
Groups with negative contributions receive $w_g = 0$. The importance scores sum to one and identify the feature groups that most support the predicted root cause.

For structural groups, which correspond to transformer components (\eg attention, QKV, FFN), each $w_g$ maps to a transformer subsystem. A high score therefore identifies the subsystem whose message-passed embedding most supports the diagnosis~\citep{wolf2024keep}. For the cross-layer observation group (\eg representation drift), a high score indicates that cross-layer behavior contributes to the diagnosis. For model-wide context groups (\eg training dynamics, validation performance), a high score means that training or validation behavior helped distinguish the predicted root cause. These non-structural groups provide supporting evidence. They should not be interpreted as the faulty component.

\section{Evaluation}
\label{sec:defaultpp_experiment}

We evaluate DEFault++ on the benchmark (\ie DEFault-bench) to assess its effectiveness across fault detection, fault categorization, root-cause diagnosis, and feature-group evidence. We report encoder and decoder results separately because the two architectures use different feature groups and label spaces.

\subsection{Evaluation Metrics}
\label{sec:defaultpp_metrics}

We evaluate DEFault++ using precision, recall, and F1. At Level~1, the faulty and correct classes are balanced by construction, so we report binary precision, recall, and F1. At Levels~2 and~3, fault categories and root causes are imbalanced, so we report macro-averaged precision, recall, and F1 across classes~\citep{naidu2023review}. We do not report accuracy because it can be misleading under class imbalance~\citep{buda2018systematic,naidu2023review}. We define true positives according to the hierarchy. Level~1 requires correct fault detection. Level~2 requires the correct fault category. Level~3 counts a prediction as correct only when both the predicted category and the predicted root cause are correct.

\subsection{Baselines}
\label{sec:defaultpp_baselines}

We compare DEFault++ with four prior fault-detection techniques for DNNs, namely AutoTrainer~\citep{autotrainer}, DeepDiagnosis~\citep{deepdiagnosis}, DeepFD~\citep{deepfd}, and DEFault~\citep{default}. These baselines distinguish faulty runs from correct runs. They do not define fault labels that match our transformer fault categories or root causes, so we limit the comparison to the fault-detection level. For the rule-based baselines, AutoTrainer and DeepDiagnosis, we use their published rules and thresholds and map the required observables to the closest available trace features in DEFault-bench. AutoTrainer's gradient-threshold rules map to the Training dynamics group, and DeepDiagnosis includes rules that partially overlap with the Attention and Score groups. For the learning-based baselines, DeepFD and DEFault, we train the original model types from their replication packages on the fault-detection task~\citep{deepfd,default}.

\subsection{Implementation}
\label{sec:defaultpp_implementation}

We implement DEFault++ in PyTorch~\citep{paszke2019pytorch}. Each feature group is encoded by a separate MLP with hidden dimension $h = 32$, the shared projection maps the concatenated group embeddings to $e = 64$ dimensions, and DEFault++ applies three rounds of FPG message passing over the group-level adjacency matrix. Training uses Adam~\citep{kingma2015adam} with learning rate $10^{-3}$, weight decay $10^{-4}$, and batch size 256. We train for up to 150 epochs with early stopping on the validation split. We set the contrastive temperature to $\tau = 0.1$ and tune the loss weights by grid search. We address fault-category and root-cause imbalance using inverse-frequency weighting in the corresponding classification losses. We use a fixed set of random seeds for reproducibility. The replication package contains the implementation, hyperparameter lists, and trained models~\citep{defaultpp_repo}. By approximately epoch 100, fault-detection F1 on the validation split shows little further improvement, while fault-categorization F1 continues to improve modestly until early stopping.

\subsection{Answering \texorpdfstring{RQ$\mathbf{_1}$:}{RQ1:} Effectiveness of Diagnosis}
\label{sec:defaultpp_rq1}

To answer RQ$_1$, we evaluate DEFault++ at its three levels, namely fault detection, fault categorization, and root-cause diagnosis.

\textbf{Fault Detection.} \Cref{tab:perf_level1_overall} shows that DEFault++ achieves F1 of 0.8259 for encoders and 0.9094 for decoders. Recall is also high on both architectures, indicating that most faulty instances are detected before the later diagnosis levels.

\begin{table}[htbp]
\centering
\caption{Overall DEFault++ performance across three levels}
\label{tab:overall_performance}

\begin{subtable}[t]{0.30\linewidth}
\centering
\caption{Level~1: Fault Detection}
\label{tab:perf_level1_overall}
\footnotesize
\setlength{\tabcolsep}{5pt}
\renewcommand{\arraystretch}{1.00}
\begin{tabular}{@{}l c c@{}}
\toprule
\textbf{Metric} & \textbf{Encoder} & \textbf{Decoder} \\
\midrule
F1          & 0.8259 & 0.9094 \\
Precision   & 0.8014 & 0.9125 \\
Recall      & 0.8520 & 0.9064 \\
\bottomrule
\end{tabular}
\end{subtable}
\hfill
\begin{subtable}[t]{0.30\linewidth}
\centering
\caption{Level~2: Fault Categorization}
\label{tab:perf_level2_overall}
\footnotesize
\setlength{\tabcolsep}{5pt}
\renewcommand{\arraystretch}{1.00}
\begin{tabular}{@{}l c c@{}}
\toprule
\textbf{Metric} & \textbf{Encoder} & \textbf{Decoder} \\
\midrule
F1          & 0.8497 & 0.8679 \\
Precision   & 0.8662 & 0.8794 \\
Recall      & 0.8403 & 0.8649 \\
\bottomrule
\end{tabular}
\end{subtable}
\hfill
\begin{subtable}[t]{0.30\linewidth}
\centering
\caption{Level~3: Root-Cause Diagnosis}
\label{tab:perf_level3_overall}
\footnotesize
\setlength{\tabcolsep}{5pt}
\renewcommand{\arraystretch}{1.00}
\begin{tabular}{@{}l c c@{}}
\toprule
\textbf{Metric} & \textbf{Encoder} & \textbf{Decoder} \\
\midrule
F1          & 0.8512 & 0.8670 \\
Precision   & 0.9407 & 0.9475 \\
Recall      & 0.7932 & 0.8225 \\
\bottomrule
\end{tabular}
\end{subtable}
\end{table}

\textbf{Fault Categorization.} \Cref{tab:perf_level2_overall} shows strong fault-categorization performance for both encoders (F1 of 0.8497) and decoders (F1 of 0.8679). Precision and recall stay close in value, which indicates that the result does not come from a strong precision-recall tradeoff. The per-category results in \cref{tab:perf_level2_perclass} show that Residual and Positional faults are among the strongest categories in both settings.

For encoder models, DEFault++ scores lowest on Variant (0.741) and Kernel (0.771). Variant differs sharply across the two families, reaching 0.948 on decoders despite a similar number of instances (25 for encoders and 23 for decoders, \cref{tab:benchmark_summary}), so the small Variant set does not by itself explain the encoder result. One possible reason is that decoder Variant faults produce cache-related differences that the encoder-only models cannot show. Kernel faults are difficult to categorize because they affect low-level execution behavior, such as numerical precision and backend selection, rather than a specific transformer component~\citep{flashattention-issue1321}.

For decoder models, DEFault++ scores lowest on FFN and KV Cache faults. FFN faults are harder to categorize because their effects can appear in downstream components, such as the residual stream and LayerNorm~\citep{vaswani2017attention}. KV Cache faults are challenging because the same cache fault can produce different measured behavior across token-generation steps~\citep{zhang2024layer, adnan2024keyformer}.

\begin{table}[htbp]
\centering
\caption{DEFault++ performance for Level~2 and Level~3 diagnosis (fault-category-wise)}
\label{tab:perclass_performance}

\begin{subtable}[t]{0.49\linewidth}
\centering
\caption{Level~2: Fault Categorization (per-category)}
\label{tab:perf_level2_perclass}
\footnotesize
\setlength{\tabcolsep}{4pt}
\renewcommand{\arraystretch}{1.00}
\begin{tabular}{@{}l c c c c c c@{}}
\toprule
& \multicolumn{3}{c}{\textbf{Encoder}} & \multicolumn{3}{c}{\textbf{Decoder}} \\
\cmidrule(lr){2-4} \cmidrule(lr){5-7}
\textbf{Category} & \textbf{F1} & \textbf{Prec.} & \textbf{Rec.} & \textbf{F1} & \textbf{Prec.} & \textbf{Rec.} \\
\midrule
Embedding   & 0.838 & 0.852 & 0.829 & 0.862 & 0.833 & 0.897 \\
FFN         & 0.836 & 0.807 & 0.871 & 0.718 & 0.673 & 0.777 \\
Kernel      & 0.771 & 0.814 & 0.740 & 0.765 & 0.848 & 0.700 \\
KV Cache    & n/a   & n/a   & n/a   & 0.758 & 0.716 & 0.822 \\
LayerNorm   & 0.790 & 0.775 & 0.820 & 0.843 & 0.937 & 0.766 \\
Masking     & 0.917 & 0.952 & 0.886 & 0.890 & 0.927 & 0.857 \\
Output      & 0.889 & 0.927 & 0.864 & 0.975 & 0.980 & 0.969 \\
Positional  & 0.919 & 0.950 & 0.891 & 0.927 & 0.949 & 0.911 \\
QKV         & 0.825 & 0.863 & 0.792 & 0.789 & 0.772 & 0.815 \\
Residual    & 0.961 & 0.965 & 0.958 & 0.995 & 0.995 & 0.995 \\
Score       & 0.861 & 0.847 & 0.876 & 0.946 & 0.937 & 0.957 \\
Variant     & 0.741 & 0.776 & 0.717 & 0.948 & 0.986 & 0.914 \\
\midrule
\textbf{Avg.} & 0.850 & 0.866 & 0.840 & 0.868 & 0.879 & 0.865 \\
\bottomrule
\end{tabular}
\end{subtable}
\hfill
\begin{subtable}[t]{0.49\linewidth}
\centering
\caption{Level~3: Root-Cause Diagnosis (per-category)}
\label{tab:perf_level3_category}
\footnotesize
\setlength{\tabcolsep}{4pt}
\renewcommand{\arraystretch}{1.00}
\begin{tabular}{@{}l c c c c c c@{}}
\toprule
& \multicolumn{3}{c}{\textbf{Encoder}} & \multicolumn{3}{c}{\textbf{Decoder}} \\
\cmidrule(lr){2-4} \cmidrule(lr){5-7}
\textbf{Category} & \textbf{F1} & \textbf{Prec.} & \textbf{Rec.} & \textbf{F1} & \textbf{Prec.} & \textbf{Rec.} \\
\midrule
Embedding   & 0.879 & 0.972 & 0.806 & 0.877 & 0.972 & 0.816 \\
FFN         & 0.804 & 0.847 & 0.778 & 0.883 & 1.000 & 0.803 \\
Kernel      & 0.814 & 1.000 & 0.706 & 0.747 & 0.988 & 0.676 \\
KV Cache    & n/a   & n/a   & n/a   & 0.881 & 1.000 & 0.817 \\
LayerNorm   & 0.821 & 0.917 & 0.749 & 0.890 & 0.969 & 0.839 \\
Masking     & 0.925 & 0.986 & 0.874 & 0.845 & 0.902 & 0.801 \\
Output      & 0.882 & 0.958 & 0.822 & 0.972 & 1.000 & 0.947 \\
Positional  & 0.930 & 0.992 & 0.886 & 0.975 & 1.000 & 0.953 \\
QKV         & 0.765 & 0.868 & 0.694 & 0.646 & 0.727 & 0.592 \\
Residual    & 0.908 & 0.927 & 0.895 & 0.958 & 0.961 & 0.955 \\
Score       & 0.835 & 0.879 & 0.799 & 0.908 & 0.926 & 0.897 \\
Variant     & 0.799 & 1.000 & 0.717 & 0.824 & 0.925 & 0.773 \\
\midrule
\textbf{Avg.} & 0.851 & 0.941 & 0.793 & 0.867 & 0.948 & 0.823 \\
\bottomrule
\end{tabular}
\end{subtable}
\end{table}

\textbf{Root-cause Diagnosis.} \Cref{tab:perf_level3_overall} shows that DEFault++ achieves F1 of 0.851 on encoders and 0.867 on decoders for overall root-cause diagnosis. We observe higher precision than recall on both architectures, indicating that incorrect root-cause predictions are less common than missed root-cause cases. The per-category results (\cref{tab:perf_level3_category}) show that DEFault++ diagnoses root causes most accurately for Positional and Residual faults. DEFault++ performs lowest when diagnosing root causes within the QKV fault category. Several QKV root causes can change projection alignment, attention scores, and update behavior in similar ways~\citep{ABNN_sigma}. This overlap may explain why QKV root causes are harder to distinguish. The Level~3 per-category results also indicate that the number of root causes within a fault category alone does not explain the performance. The Residual fault category includes five root causes and remains one of the strongest categories for root-cause diagnosis. QKV includes four root causes but remains the weakest. This pattern suggests that separability within each category plays a larger role than the number of root causes alone.

\begin{rqsummary}
\textbf{Summary of RQ$\mathbf{_1}$:} DEFault++ diagnoses faults at all three levels on the evaluated encoder and decoder models, with F1 of 0.826--0.909 for detection, 0.850--0.868 for categorization, and 0.851--0.867 for root-cause diagnosis. Decoders score slightly higher than encoders at every level, and Level~3 precision exceeds 0.94 on both.
\end{rqsummary}

\subsection{Answering {\texorpdfstring{RQ$\mathbf{_2}$}{RQ2}:} Comparison with Existing Techniques}
\label{sec:defaultpp_rq2}

To answer RQ$_2$, we compare DEFault++ with four existing DNN fault-detection techniques. AutoTrainer~\citep{autotrainer} and DeepDiagnosis~\citep{deepdiagnosis} report training-symptom outputs. DeepFD~\citep{deepfd} and DEFault~\citep{default} use DNN-level fault categories, which do not correspond to the transformer component categories or root causes used in our evaluation. We therefore restrict the comparison to fault detection (Level~1).

The baseline results (\Cref{tab:rq2_detection_comparison}) highlight the difference between general DNN fault detection and transformer-component fault detection. AutoTrainer and DeepDiagnosis use fixed training symptoms (\eg gradient failures, loss oscillation, overfitting). Many DEFault-bench faults do not necessarily produce these training-level symptoms. A masking, QKV, positional, or cache fault can change internal transformer behavior while the loss curve and numerical values remain within ordinary ranges. As a result, rule-based methods can miss faults that do not match their predefined symptoms. In contrast, DeepFD and DEFault perform better than the rule-based methods because they learn from runtime features rather than relying on fixed rules. However, they still focus on general DNN training behavior and do not directly capture transformer-specific behavior (\eg attention distributions, QKV alignment, cache behavior, or component-level propagation). These results suggest that transformer fault detection is not only a model-level prediction task. It also requires component-level measurements and the dependencies among transformer components.

\begin{table}[htbp]
\centering
\caption{Fault-detection comparison between DEFault++ and baselines (F1)}
\label{tab:rq2_detection_comparison}
\small
\begin{adjustbox}{max width=\textwidth}
\begin{tabular}{l l c c}
\toprule
\textbf{Technique} & \textbf{Type} & \textbf{Encoder} & \textbf{Decoder} \\
\midrule
\rowcolor{oursrow}
\textbf{DEFault++} & Neural hierarchical & $\mathbf{0.8259}$ & $\mathbf{0.9094}$ \\
DEFault~\citep{default} & Hierarchical RF & $0.5660$ & $0.7020$ \\
DeepFD~\citep{deepfd} & Flat ensemble & $0.5140$ & $0.6550$ \\
DeepDiagnosis~\citep{deepdiagnosis} & Rule-based (8 rules) & $0.3380$ & $0.4820$ \\
AutoTrainer~\citep{autotrainer} & Rule-based (5 rules) & $0.2840$ & $0.4270$ \\
\bottomrule
\end{tabular}
\end{adjustbox}
\end{table}

\begin{rqsummary}
\textbf{Summary of RQ$\mathbf{_2}$:} DEFault++ outperforms all four baselines in detecting faults in transformer models, none of which is designed to capture transformer-component behavior. Our findings suggest that many transformer faults may not be captured by standard runtime metrics alone. Detecting these faults benefits from transformer-specific measurements and architecture-aware diagnosis.
\end{rqsummary}

\subsection{Answering {RQ\texorpdfstring{$\mathbf{_3}$}{3}:} Ablation}
\label{sec:defaultpp_rq3}

To answer RQ$_3$, we ablate two major parts of DEFault++ and assess their impact on performance. In particular, we focus on (a) message passing in the Fault Propagation Graph (FPG) and (b) the separation loss $\mathcal{L}_{\text{sep}}$. FPG message passing applies to all three diagnosis levels, while $\mathcal{L}_{\text{sep}}$ applies only to root-cause diagnosis.

We evaluate four model variants. The full DEFault++ model uses both FPG message passing and separation loss. The $-$FPG variant removes message passing and keeps the remaining model unchanged. The $-$Sep variant removes the separation loss by setting $\beta = 0$ and $\gamma = 0$. The $-$FPG $-$Sep variant removes both components and keeps only the feature-group encoders, shared projection, and prediction heads.

We also evaluate two graph-topology variants to separate the effect of the FPG's transformer-specific structure from the influence of graph connectivity alone. The \emph{Rewired variant} keeps the same number of nodes and edges as the FPG but randomly reassigns edge endpoints while preserving the degree distribution~\citep{maslov2002specificity}. The \emph{Random variant} replaces the FPG with an Erd\H{o}s--R\'{e}nyi graph~\citep{erdos1959random} with approximately the same edge density.

\textbf{Fault Detection \& Categorization.} For Level~1 fault detection, removing FPG message passing shows a slight performance decline on both architectures (\cref{tab:abl_l1}). This result suggests that the feature groups already contain enough information for binary fault detection, while FPG message passing adds a slight advantage by combining evidence from related components.

For Level~2 fault categorization, removing FPG message passing leads to a larger performance drop (\cref{tab:abl_l2}). Categorization requires the model to distinguish fault types that can affect related parts of the transformer. For example, QKV and Score faults both affect attention behavior, but they reach downstream components through different propagation paths. FPG message passing allows the model to combine evidence from related feature groups before predicting the fault category.

\begin{table}[htbp]
\centering
\small
\caption{Ablation studies for Level~1 fault detection and Level~2 fault categorization.}
\begin{subtable}[t]{0.48\linewidth}
\centering
\caption{Level~1 fault detection}
\label{tab:abl_l1}
\begin{tabular}{l l c c c}
\toprule
\textbf{Arch.} & \textbf{Variant} & \textbf{F1} & \textbf{Prec.} & \textbf{Rec.} \\
\midrule
Encoder & DEFault++ & 0.826 & 0.801 & 0.852 \\
        & $-$FPG     & 0.782 & 0.752 & 0.816 \\
\midrule
Decoder & DEFault++ & 0.909 & 0.912 & 0.906 \\
        & $-$FPG     & 0.890 & 0.890 & 0.891 \\
\bottomrule
\end{tabular}
\end{subtable}\hfill
\begin{subtable}[t]{0.48\linewidth}
\centering
\caption{Level~2 fault categorization}
\label{tab:abl_l2}
\begin{tabular}{l l c c c}
\toprule
\textbf{Arch.} & \textbf{Variant} & \textbf{F1} & \textbf{Prec.} & \textbf{Rec.} \\
\midrule
Encoder & DEFault++ & 0.850 & 0.866 & 0.840 \\
        & $-$FPG     & 0.742 & 0.757 & 0.731 \\
\midrule
Decoder & DEFault++ & 0.868 & 0.879 & 0.865 \\
        & $-$FPG     & 0.788 & 0.797 & 0.783 \\
\bottomrule
\end{tabular}
\end{subtable}
\end{table}

\textbf{Root-cause Diagnosis.} At Level~3, removing the separation loss causes a larger performance drop than removing FPG message passing (\cref{tab:abl_l3}). This pattern suggests that the main challenge at this level is distinguishing root causes within the same fault category. Root causes in the same category often affect the same transformer component, which can make their runtime measurements similar. The separation loss directly targets this problem by encouraging embeddings from different root causes in the same category to move apart.

FPG message passing also contributes to Level~3, but its effect is smaller than that of the separation loss. Removing both FPG message passing and the separation loss leads to the lowest Level~3 performance, suggesting that they provide complementary benefits.

\begin{table}[htbp]
\centering
\caption{Ablation study of Level~3 root cause diagnosis}
\label{tab:abl_l3}
\small
\begin{tabular}{l l c c c}
\toprule
\textbf{Architecture} & \textbf{Variant} & \textbf{F1} & \textbf{Precision} & \textbf{Recall} \\
\midrule
Encoder & DEFault++ & 0.851 & 0.941 & 0.793 \\
        & $-$FPG & 0.812 & 0.933 & 0.776 \\
        & $-$Sep & 0.756 & 0.910 & 0.752 \\
        & $-$FPG$ -$Sep & 0.701 & 0.883 & 0.690 \\
\midrule
Decoder & DEFault++ & 0.867 & 0.948 & 0.823 \\
        & $-$FPG & 0.833 & 0.940 & 0.809 \\
        & $-$Sep & 0.781 & 0.925 & 0.781 \\
        & $-$FPG$ -$Sep & 0.734 & 0.896 & 0.721 \\
\bottomrule
\end{tabular}
\end{table}

\textbf{Topology Ablation.} \Cref{tab:abl_topology} shows that the original FPG gives the best performance at every diagnosis level for both encoders and decoders. The Rewired and Random variants degrade performance, and removing FPG message passing gives the lowest performance. The drop from the original FPG to the Rewired and Random variants shows that transformer-specific propagation links matter beyond graph connectivity alone. We observe that the topology effect is largest at the fault-categorization level. The FPG-removal ablation shows the same Level~2 pattern, with the largest performance drop appearing in fault categorization (\cref{tab:abl_l2}). Since fault categorization depends more on propagation structure, it must differentiate fault categories that can affect related transformer components. The topology differences are smaller at Level~1 and Level~3. Level~1 is a binary detection task, while Level~3 depends more on within-category root-cause separation through $\mathcal{L}_{\text{sep}}$. These ablations show that the FPG structure helps categorization, but they do not establish that its edges match the trained model's actual fault-propagation paths.

\begin{table}[htbp]
\centering
\caption{Topology ablation results using F1 across graph variants}
\label{tab:abl_topology}
\footnotesize
\begin{tabular}{@{} l l c c c c @{}}
\toprule
\textbf{Level} & \textbf{Architecture} & \textbf{DEFault++} & \textbf{Rewired} & \textbf{Random} & \textbf{$-$FPG} \\
\midrule
L1 & Encoder & 0.826 & 0.790 & 0.786 & 0.782 \\
   & Decoder & 0.909 & 0.894 & 0.892 & 0.890 \\
\midrule
L2 & Encoder & 0.850 & 0.770 & 0.754 & 0.742 \\
   & Decoder & 0.868 & 0.807 & 0.796 & 0.788 \\
\midrule
L3 & Encoder & 0.851 & 0.820 & 0.816 & 0.812 \\
   & Decoder & 0.867 & 0.838 & 0.835 & 0.833 \\
\bottomrule
\end{tabular}
\end{table}

\begin{rqsummary}
\textbf{Summary of RQ$\mathbf{_3}$:} Message passing in the Fault Propagation Graph (FPG) contributes most to DEFault++'s fault-categorization performance, while the separation loss contributes most to root-cause diagnosis. The topology ablation shows that the defined FPG structure is more effective than rewired or random graph variants, especially for fault categorization.
\end{rqsummary}

\subsection{Answering {RQ\texorpdfstring{$\mathbf{_4}$}{RQ4}:} Explanation Faithfulness}
\label{sec:defaultpp_rq4}

DEFault++ uses the prototype classifier for root-cause diagnosis. The prototype distance decomposes over feature groups, so DEFault++ can measure how much each group contributes to the selected root-cause prototype. To answer RQ$_4$, we assess the faithfulness of these feature-group explanations to DEFault++'s prototype-based root-cause decisions.

\begin{figure}[htbp]
\centering
\includegraphics[width=\linewidth]{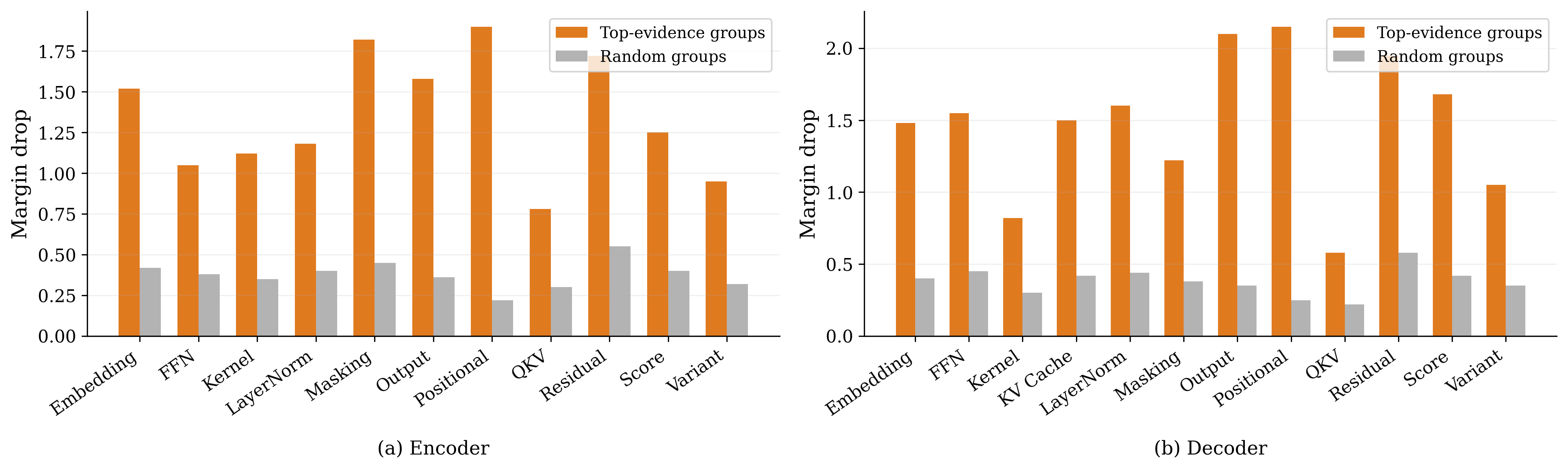}
\caption{Group ablation analysis for the FPG-based explanation}
\Description{Visualization of group ablation analysis for the FPG-based explanation.}
\label{fig:rq4_group_ablation}
\end{figure}

We first evaluate the reported feature groups through ablation. For each prediction, we remove the two feature groups that DEFault++ ranks highest. We then recompute the prototype-distance gap between the predicted root-cause prototype and the nearest alternative prototype. The gap is computed as the distance to the nearest alternative prototype minus the distance to the predicted prototype. Larger gaps indicate stronger support for the predicted root cause. As a baseline, we repeat the same test after removing two randomly selected feature groups. \Cref{fig:rq4_group_ablation} shows that removing the top-ranked groups reduces the prototype-distance gap more than removing random groups. A similar pattern appears for both encoder and decoder models, and it applies to all fault categories.

\begin{figure}[htbp]
\centering
\includegraphics[width=\linewidth]{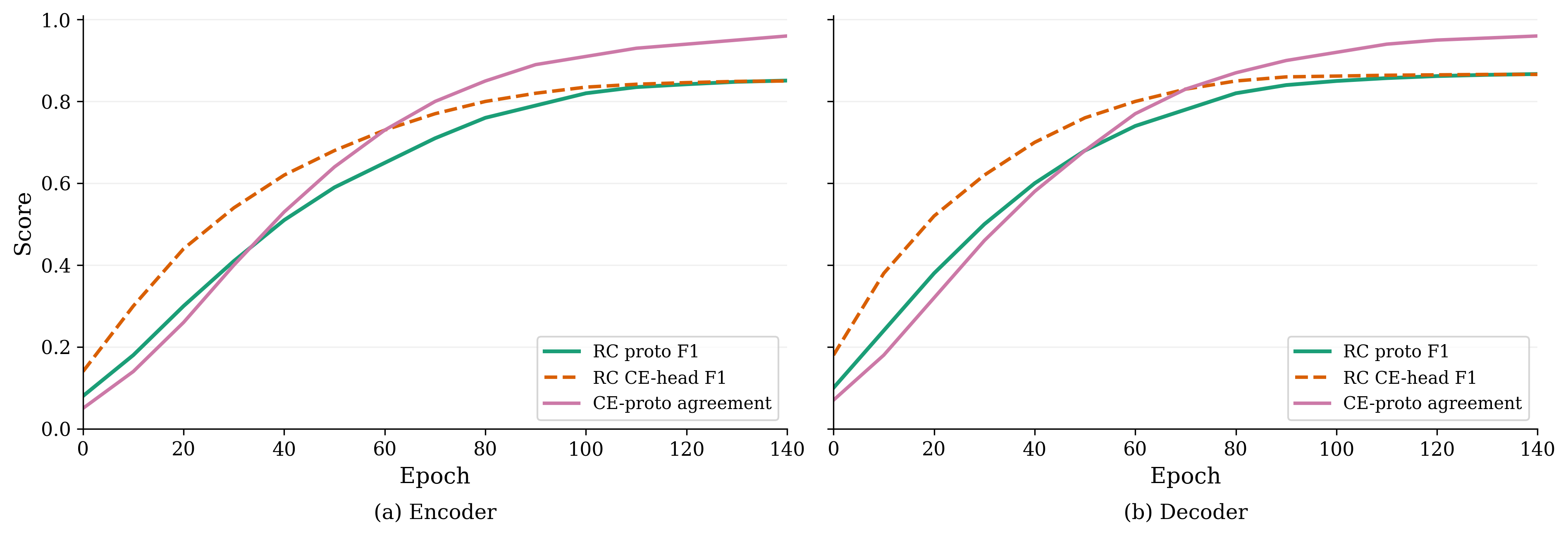}
\caption{Prototype-label agreement during training}
\Description{A plot showing prototype-label agreement during training.}
\label{fig:defaultpp_prototype_faithfulness}
\end{figure}

We next evaluate the prototype classifier used for the explanation. DEFault++ also trains a supervised root-cause classifier, which predicts root causes directly from the training labels. We compare these two classifiers by measuring how often they predict the same root cause for the same sample. \Cref{fig:defaultpp_prototype_faithfulness} shows that prototype accuracy increases during training, and its agreement with the supervised classifier also increases for both encoder and decoder models. Since DEFault++ computes explanations from the prototype classifier, the increasing agreement supports the use of prototype-based explanations for the learned root-cause decisions.

DEFault-bench provides the root-cause annotation for each faulty run, but it does not specify which feature groups should explain that fault. We therefore assess faithfulness by testing whether the reported groups affect DEFault++'s prototype-based decision. This evaluation supports internal faithfulness, but it does not establish the reported groups as the true causal explanation for the fault.

\begin{rqsummary}
\textbf{Summary of RQ$\mathbf{_4}$:} DEFault++ identifies the feature groups that support its root-cause diagnoses. Our results show that higher-ranked feature groups provide stronger diagnostic evidence than randomly selected groups, and that the explanation mechanism remains aligned with the model's final prediction.
\end{rqsummary}

\section{Real-world Fault Evaluation}
\label{sec:defaultpp_casestudy}

We use real-world GitHub issues to evaluate whether DEFault++ generalizes beyond the synthetic faults in DEFault-bench. Following prior work~\citep{deeplocalize, fl4deep, ABNN_sigma}, we collect 20 transformer-related bug-fix issues from open-source repositories and reproduce 11 of them. For each reproduced issue, we compare the faulty version with the fixed version using the same feature extraction procedure as in the main evaluation. DEFault++ detects 8 of the 11 faults, assigns the correct fault category for 8, and identifies the correct root cause for 4 (\cref{tab:realworld_eval}). Fault detection (FD) checks whether DEFault++ marks the faulty version as faulty. Fault categorization (FC) checks whether it assigns the fault to the correct transformer category. Root-cause diagnosis (RC) checks whether the predicted cause matches the cause confirmed by the corresponding bug fix.

\begin{table}[htbp]
\caption{Real-world evaluation of DEFault++ on 11 transformer faults. FD, FC, and RC denote fault detection, fault categorization, and root-cause diagnosis. \textcolor{markred}{Crash*} marks a run that fails before feature extraction, which we count as not applicable rather than as a diagnostic failure.}
\label{tab:realworld_eval}
\renewcommand{\arraystretch}{1.0}
\setlength{\tabcolsep}{3.5pt}
\footnotesize
\begin{tabularx}{\linewidth}{@{}p{2.0cm} p{2.0cm} c c c p{3.8cm} X@{}}
\toprule
\textbf{Source} &
\textbf{Fault category} &
\multicolumn{3}{c}{\textbf{DEFault++ result}} &
\textbf{Feature Importance (top)} &
\textbf{Relevant Scope of DEFault++} \\
\cmidrule(lr){3-5}
& & \textbf{FD} & \textbf{FC} & \textbf{RC} & & \\
\midrule

\makecell[l]{jax\#23349}
& Masking
& \cmark & \cmark & \cmark
& Attention: padding attention mass
& Attention feature group \\

\makecell[l]{pytorch\#103082}
& Masking
& \cmark & \cmark & \cmark
& Attention: future attention mass
& Attention feature group \\

\makecell[l]{transformers\#19045}
& Positional
& \cmark & \cmark & \xmark
& Positional: positional sensitivity
& Forward sequential propagation (M1) \\

\makecell[l]{transformers\#17886}
& Positional
& \cmark & \cmark & \xmark
& Attention: inter-head cosine similarity
& Forward sequential propagation (M1) \\

\makecell[l]{ai-edge-torch\#6}
& QKV
& \cmark & \cmark & \cmark
& Score: pre-softmax score
& Simultaneous propagation (M2) \\

\makecell[l]{nsa-pytorch\#20}
& KV Cache
& \cmark & \cmark & \xmark
& KV Cache: cache-related behavior
& Generation-time cache update path \\

\makecell[l]{transformers\#37574}
& KV Cache
& \xmark & \xmark & \xmark
& --
& Generation-time cache behavior \\

\makecell[l]{transformers\#36096}
& Score
& \multicolumn{3}{c}{\textcolor{markred}{Crash*}}
& --
& Runtime failure before feature extraction \\

\makecell[l]{pytorch\#116333}
& Kernel
& \xmark & \xmark & \xmark
& --
& Backend stride validation \\

\makecell[l]{transformers\#35896}
& Variant
& \cmark & \cmark & \xmark
& Attention: attention entropy
& Cross-layer propagation (M4) \\

\makecell[l]{diffusers\#11903}
& QKV
& \cmark & \cmark & \cmark
& Optimization: projection update ratio (Dynamic parameter registration)
& QKV projection/update path \\

\midrule
\multicolumn{7}{@{}l}{Summary: FD = 8/11, FC = 8/11, RC = 4/11.} \\
\bottomrule
\end{tabularx}
\end{table}

For the real-world faults that DEFault++ detects and categorizes correctly, the top-ranked feature groups generally align with the reported fault category. For the masking faults, DEFault++ assigns high importance to attention-mass features, which measure attention given to invalid tokens. For the QKV faults, the top-ranked groups include score and projection-update features, which correspond to QKV projection behavior and attention-score computation. For positional and variant faults, DEFault++ assigns high importance to position-sensitive and attention-behavior features, although these cases are not always diagnosed at the root-cause level. DEFault++ detects and categorizes several real-world faults correctly but does not always identify the correct root cause, as in the positional, KV Cache, and variant cases. This matches our benchmark results, which show that root-cause diagnosis is the most difficult level.

DEFault++ does not identify three real-world faults. In \texttt{transformers}\#37574, the fault appears during autoregressive generation, where the model stores KV states from earlier decoding steps and reuses them when generating later tokens. DEFault++ extracts cache-related behavior during fine-tuning, but it does not trace how the cache changes at each generation step, which can limit its ability to diagnose faults that appear only through generation-time cache updates. In \texttt{pytorch}\#116333, the fault depends on backend stride validation rather than transformer-layer behavior, so diagnosing it would require system-level information beyond the model trace DEFault++ uses. Finally, \texttt{transformers}\#36096 falls outside DEFault++'s scope because the fault crashes before a valid trace can be collected.

\begin{figure}[htbp]
\centering
\begin{subfigure}[t]{0.50\linewidth}
\caption{Hierarchical diagnostic output}
\label{fig:defaultpp_case_qkv_a}
\centering
\begin{tikzpicture}[
    arrow/.style={-{Stealth[length=4pt]}, thick, black!50},
]

\def\boxleft{0.0}
\def\boxright{7.2}
\def\boxmid{3.6}
\def\labelboxwidth{2.6}
\def\textstart{0.2}

\draw[black!50, rounded corners=3pt] (\boxleft, 0.0) rectangle (\boxright, -1.3);
\fill[rqblue, rounded corners=2pt] (\boxmid-\labelboxwidth/2, -0.12) rectangle (\boxmid+\labelboxwidth/2, -0.57);
\node[text=white, font=\small] at (\boxmid, -0.345) {Fault Detection};
\node[anchor=center, align=center, font=\small] at (\boxmid, -0.85) {\textbf{Faulty} \quad (confidence: 0.98)};
\node[anchor=center, align=center, font=\scriptsize, text=black!60, text width=6.9cm] at (\boxmid, -1.10) {Normal training loss, abnormal projection-update behavior};

\draw[arrow] (\boxmid, -1.3) -- (\boxmid, -1.5);

\draw[black!50, rounded corners=3pt] (\boxleft, -1.5) rectangle (\boxright, -2.8);
\fill[rqblue, rounded corners=2pt] (\boxmid-\labelboxwidth/2, -1.62) rectangle (\boxmid+\labelboxwidth/2, -2.07);
\node[text=white, font=\small] at (\boxmid, -1.845) {Fault Categorization};
\node[anchor=center, align=center, font=\small] at (\boxmid, -2.35) {\textbf{QKV Fault} \quad (confidence: 0.91)};
\node[anchor=center, align=center, font=\scriptsize, text=black!60, text width=6.9cm] at (\boxmid, -2.60) {Deviations in QKV alignment, projection update, attention entropy};

\draw[arrow] (\boxmid, -2.8) -- (\boxmid, -3.0);

\draw[black!50, rounded corners=3pt] (\boxleft, -3.0) rectangle (\boxright, -4.55);
\fill[rqblue, rounded corners=2pt] (\boxmid-\labelboxwidth/2, -3.12) rectangle (\boxmid+\labelboxwidth/2, -3.57);
\node[text=white, font=\small] at (\boxmid, -3.345) {Root-Cause};
\node[anchor=center, align=center, font=\small] at (\boxmid, -3.95) {\textbf{Dynamic Parameter Registration} {\scriptsize(confidence: 0.88)}};
\node[anchor=center, align=center, font=\scriptsize, text=black!60, text width=6.9cm] at (\boxmid, -4.30) {Stale projection update path: near-zero update ratio for \texttt{q\_proj}/\texttt{v\_proj} while the fused path is active};

\end{tikzpicture}
\end{subfigure}\hfill
\begin{subfigure}[t]{0.48\linewidth}
\caption{Feature-group importance ($w_g$)}
\label{fig:defaultpp_case_qkv_b}
\centering
\begin{tikzpicture}[
    barfill/.style={fill=rqblue, draw=rqblue!80},
    barfilllight/.style={fill=rqblue!25, draw=rqblue!40},
]

\def\barstart{2.2}
\def\barscale{4.5}
\def\barheight{0.30}
\def\bartextoffset{0.12}
\def\rowqkv{-0.6}
\def\rowopt{-1.2}
\def\rowatt{-1.8}
\def\rowscore{-2.4}
\def\rowother{-3.0}

\node at (0, 0.0) {};

\node[anchor=east, font=\small] at (\barstart - 0.10, \rowqkv) {QKV alignment};
\fill[barfill] (\barstart, \rowqkv-\barheight/2) rectangle (\barstart+0.45*\barscale, \rowqkv+\barheight/2);
\node[anchor=west, font=\small\bfseries] at (\barstart+0.45*\barscale+\bartextoffset, \rowqkv) {45\%};

\node[anchor=east, font=\small] at (\barstart - 0.10, \rowopt) {Optimization};
\fill[barfill] (\barstart, \rowopt-\barheight/2) rectangle (\barstart+0.28*\barscale, \rowopt+\barheight/2);
\node[anchor=west, font=\small\bfseries] at (\barstart+0.28*\barscale+\bartextoffset, \rowopt) {28\%};

\node[anchor=east, font=\small] at (\barstart - 0.10, \rowatt) {Attention};
\fill[barfilllight] (\barstart, \rowatt-\barheight/2) rectangle (\barstart+0.12*\barscale, \rowatt+\barheight/2);
\node[anchor=west, font=\small] at (\barstart+0.12*\barscale+\bartextoffset, \rowatt) {12\%};

\node[anchor=east, font=\small] at (\barstart - 0.10, \rowscore) {Score};
\fill[barfilllight] (\barstart, \rowscore-\barheight/2) rectangle (\barstart+0.07*\barscale, \rowscore+\barheight/2);
\node[anchor=west, font=\small] at (\barstart+0.07*\barscale+\bartextoffset, \rowscore) {7\%};

\node[anchor=east, font=\small] at (\barstart - 0.10, \rowother) {Other};
\fill[barfilllight] (\barstart, \rowother-\barheight/2) rectangle (\barstart+0.08*\barscale, \rowother+\barheight/2);
\node[anchor=west, font=\scriptsize] at (\barstart+0.08*\barscale+\bartextoffset, \rowother) {8\%};

\draw[decorate, decoration={brace, amplitude=4pt}, thick, rqblue]
    (\barstart+0.45*\barscale+0.75, \rowqkv+\barheight/2+0.05)
    -- (\barstart+0.45*\barscale+0.75, \rowopt-\barheight/2-0.05)
    node[midway, right=4pt, font=\scriptsize, text=rqblue, align=left, text width=2.2cm]
    {73\% in QKV alignment \& optimization};
\node at (0, -4.55) {};

\end{tikzpicture}
\end{subfigure}

\Description{Two-panel figure. The left panel shows three stacked boxes for the hierarchical DEFault++ diagnosis: fault detection identifies a faulty model, fault categorization assigns the QKV category, and root cause diagnosis returns the dynamic parameter registration root cause. The right panel is a horizontal bar chart of feature-group importance, with QKV alignment at 45 percent and optimization at 28 percent highlighted as accounting for 73 percent of the total.}
\caption{DEFault++ diagnosis and feature-group importance for the stale QKV fusion fault in \cref{lst:qkv_fusion_bug}}
\label{fig:defaultpp_case_qkv}
\end{figure}

We use \texttt{diffusers}\#11903 as a real-world case study for DEFault++. The fault redirects the forward pass to a fused QKV projection while LoRA updates the original projection modules. In this case, training completes without an explicit failure, but the updates do not affect the projection path used during execution. \Cref{fig:defaultpp_case_qkv} shows the DEFault++ output for this case. DEFault++ detects the run as faulty, assigns it to the QKV category, and diagnoses the root cause as dynamic parameter registration. The explanation assigns most of the normalized feature importance to QKV Alignment and Optimization, which together account for 73\% of the score. These groups point to a mismatch between the projection modules updated by LoRA and the fused QKV projection used in the forward pass. The GitHub bug fix addresses the same mismatch by ensuring that LoRA updates the projection module executed by the model. This case also shows that DEFault++ generalizes within the dynamic parameter registration root cause rather than memorizing a single operator. The training operator \opid{QFG} freezes QKV gradients while the parameters stay in the forward path, whereas \texttt{diffusers}\#11903 is the opposite case, where gradients flow but the parameters leave the forward path. DEFault++ recognizes the second case through the same near-zero QKV update-activity signature, which suggests the root-cause label spans sub-mechanisms within a category rather than fitting one operator.

\section{Developer Study}
\label{sec:defaultpp_devstudy}
We evaluate DEFault++ through a developer study using real-world transformer debugging scenarios derived from reproduced GitHub issues. Participants review four scenarios and select the most appropriate repair action for each one. For two scenarios, participants receive only the debugging artifacts available from the reproduced issue (\eg logs, model outputs). For the other two scenarios, participants receive the same artifacts together with the diagnosis generated by DEFault++. We use the study to measure how DEFault++ affects repair correctness, confidence, and self-reported completion time, and how participants perceive the usefulness of its diagnoses.

\textbf{Developer Study Design.} We use a within-subjects, counterbalanced design with two conditions~\citep{wohlin2012experimentation}. In the baseline condition, participants receive the scenario description, observed behavior, logs, and metrics. In the DEFault++ assisted condition, they receive the same materials plus the DEFault++ diagnosis, including fault status, fault category, predicted root cause, and supporting feature group. Each participant completes all four scenarios, with two scenarios assigned to each condition. We counterbalance condition assignment across two survey forms, following prior scenario-based developer studies of DL debugging techniques~\citep{umlaut,kunit}.

The four scenarios cover Masking, Positional, KV Cache, and Variant faults (\cref{tab:devstudy_scenarios}). For S1, DEFault++ provides a correct diagnosis at all three levels. For S2, S3, and S4, DEFault++ correctly detects the fault and predicts the correct fault category, but predicts an incorrect root cause (\cref{tab:devstudy_diagnosis}). We include these imperfect-diagnosis scenarios to evaluate whether category-level diagnosis and supporting feature groups can still guide repair selection when the root-cause prediction is incorrect.

The primary outcome is repair correctness, scored against the ground-truth fix from the source issue. We also collect self-reported confidence on a 5-point scale and self-reported time as a perceived-effort measure. After completing the scenarios, participants rate the clarity, usefulness, and practical value of the DEFault++ diagnosis on 5-point Likert scales and answer preference questions comparing the two conditions.

\begin{table}[htbp]
\caption{Developer-study scenarios and correct repair actions}
\label{tab:devstudy_scenarios}
\footnotesize
\setlength{\tabcolsep}{3pt}
\renewcommand{\arraystretch}{1.0}
\begin{tabularx}{\linewidth}{@{}l l l X@{}}
\toprule
\textbf{Scenario} & \textbf{Source Issue} & \textbf{Category} & \textbf{Repair Action} \\
\midrule
S1: Cached decoding visibility
& pytorch \#103082
& Masking
& Fix the causal mask for cached decoding so that the new token attends to the correct cached key positions. \\

S2: Position-dependent scoring drift
& transformers \#19045
& Positional
& Recompute relative-position distances from the true token positions during cached decoding. \\

S3: Incremental state update mismatch
& nsa-pytorch \#20
& KV Cache
& Write the current token's key-value vectors into the cache before computing attention and prediction scores. \\

S4: Local-attention layer assignment
& transformers \#35896
& Variant
& Fix the layer-selection condition so that local/sliding-window attention is applied to the intended layers. \\
\bottomrule
\end{tabularx}
\end{table}
\begin{table}[htbp]
\caption{DEFault++ diagnosis shown to participants in each developer-study scenario}
\label{tab:devstudy_diagnosis}
\footnotesize
\setlength{\tabcolsep}{5pt}
\renewcommand{\arraystretch}{1.00}
\begin{tabular}{@{}lllll@{}}
\toprule
\textbf{Scenario} & \textbf{Level~1} & \textbf{Level~2} & \textbf{Level~3 (predicted)} & \textbf{Level~3 (ground truth)} \\
\midrule
S1 (Masking)    & Faulty~\cmark & Masking~\cmark    & Dynamic mask~\cmark       & Dynamic mask \\
S2 (Positional) & Faulty~\cmark & Positional~\cmark & Indexing~\xmark           & Relative position \\
S3 (KV Cache)   & Faulty~\cmark & KV Cache~\cmark   & Cache invalidation~\xmark & Update synchronization \\
S4 (Variant)    & Faulty~\cmark & Variant~\cmark    & Dynamic dispatch~\xmark   & Variant configuration \\
\bottomrule
\end{tabular}
\end{table}
\textbf{Developer Study Participants.} The Dalhousie University Research Ethics Board approved the study.\footnote{\url{https://www.dal.ca/research-and-innovation/support-for-researchers/responsible-conduct-research/human-ethics.html}} Participants are eligible if they have experience with transformer architectures in training, fine-tuning, inference, debugging, or evaluation~\citep{umlaut,kunit}. We recruit 21 participants through graduate-student mailing lists, academic networks, and direct outreach (\Cref{tab:devstudy_demographics}). Participation is anonymous, voluntary, and uncompensated, and runs through Microsoft Forms.

\begin{table}[htbp]
\centering
\caption{Developer study: participant demographics and repair outcomes (N\,=\,21 participants)}
\Description{Two subtables. The left subtable lists participant demographics across ML/DL experience, prior transformer debugging, and role. The right subtable reports repair accuracy and confidence for four scenarios under baseline and assisted conditions, with totals.}
\label{tab:devstudy}
\begin{subtable}[t]{0.38\linewidth}
\centering
\caption{Participant demographics}
\label{tab:devstudy_demographics}
\small
\setlength{\tabcolsep}{3pt}
\renewcommand{\arraystretch}{1.0}
\begin{tabular}{@{}l l r@{}}
\toprule
\textbf{Characteristic} & \textbf{Value} & \textbf{N} \\
\midrule
ML/DL experience & $<$1 year   & 1  \\
                 & 1--2 years  & 6  \\
                 & 3--5 years  & 10 \\
                 & $>$5 years  & 4  \\
\addlinespace
Prior transformer & Yes & 15 \\
debugging experience        & No  & 6  \\
\addlinespace
Current Role              & Research assistant   & 10 \\
(multi-select)    & Graduate student     & 11 \\
                  & Industry practitioner & 9  \\
\bottomrule
\end{tabular}
\end{subtable}\hfill
\begin{subtable}[t]{0.60\linewidth}
\centering
\caption{Repair accuracy and confidence by scenario and condition}
\label{tab:devstudy_results}
\small
\setlength{\tabcolsep}{4pt}
\renewcommand{\arraystretch}{1.0}
\begin{tabular}{@{}l l r l r r@{}}
\toprule
\textbf{Scenario} & \textbf{Condition} & \textbf{N} & \textbf{Correct} & \textbf{Accuracy} & \textbf{Confidence} \\
\midrule
S1 (Masking)    & Baseline & 12 & 8/12  & 66.7\%  & 3.75 \\
S1 (Masking)    & Assisted &  9 & 9/9   & 100.0\% & 4.11 \\
\addlinespace
S2 (Positional) & Baseline &  9 & 4/9   & 44.4\%  & 2.78 \\
S2 (Positional) & Assisted & 12 & 10/12 & 83.3\%  & 3.75 \\
\addlinespace
S3 (KV Cache)   & Baseline & 12 & 7/12  & 58.3\%  & 3.50 \\
S3 (KV Cache)   & Assisted &  9 & 8/9   & 88.9\%  & 4.11 \\
\addlinespace
S4 (Variant)    & Baseline &  9 & 5/9   & 55.6\%  & 3.33 \\
S4 (Variant)    & Assisted & 12 & 8/12  & 66.7\%  & 4.08 \\
\midrule
Total           & Baseline & 42 & 24/42 & 57.1\%  & 3.38 \\
Total           & Assisted & 42 & 35/42 & 83.3\%  & 4.00 \\
\bottomrule
\end{tabular}
\end{subtable}
\end{table}

\begin{figure}[htbp]
\centering
\includegraphics[width=\linewidth]{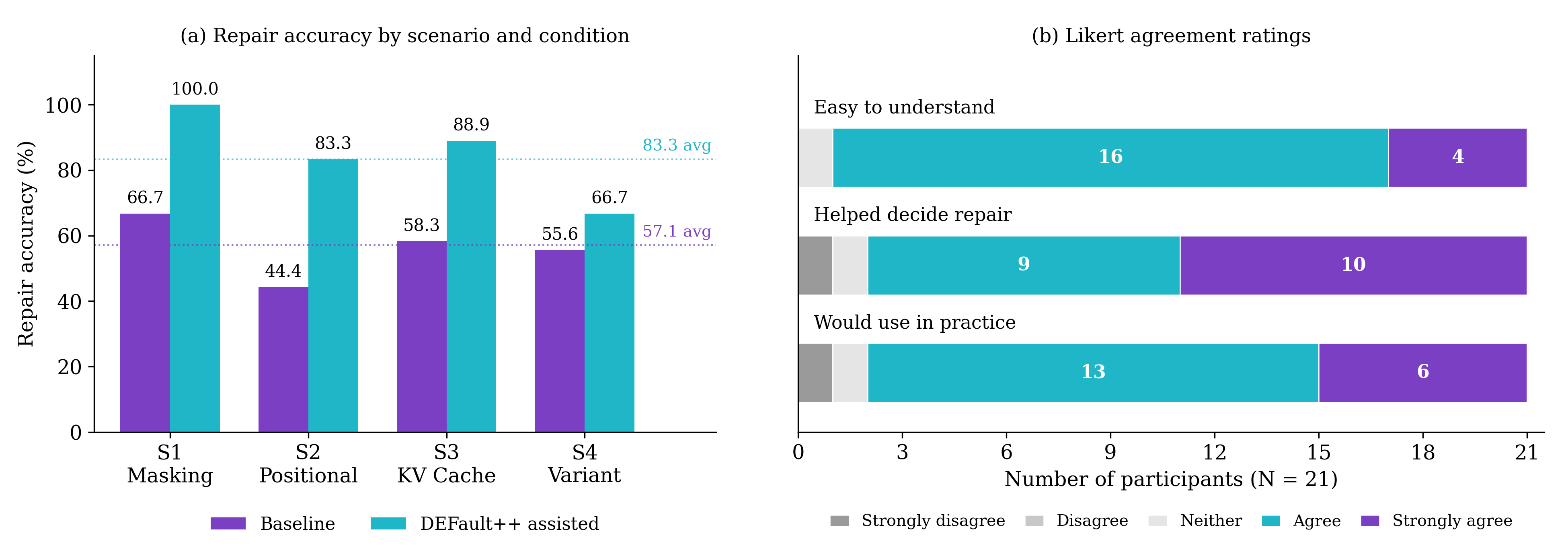}
\Description{Side-by-side panels. Panel (a) is a bar chart of repair accuracy by scenario (S1 Masking, S2 Positional, S3 KV Cache, S4 Variant) for baseline and DEFault++ assisted conditions, with horizontal lines marking 57.1\% baseline and 83.3\% assisted averages. Panel (b) is a stacked horizontal bar chart of Likert agreement ratings for three statements: easy to understand, helped decide repair, and would use in practice.}
\caption{Developer study results: (a)~Repair accuracy by scenario and condition, (b)~Likert agreement ratings}
\label{fig:devstudy_results}
\end{figure}

\textbf{Developer Study Results.} We report descriptive statistics, following prior vignette-based debugging studies such as UMLAUT~\citep{umlaut} and KUnit~\citep{kunit}. Each participant sees only two scenarios per condition, so we treat the difference between conditions as an effect-size estimate rather than as a hypothesis test.

Participants select the correct repair more often with DEFault++ assistance than in the baseline condition. Repair accuracy increases from 57.1\% in the baseline condition to 83.3\% in the assisted condition (\cref{tab:devstudy_results,fig:devstudy_results}). Confidence also increases from 3.38 to 4.00. The largest gains appear in S2 and S3, where DEFault++ predicts the correct fault category but an incorrect root cause. This suggests that category-level diagnosis and supporting feature groups can still help developers choose the correct repair direction when the root-cause prediction is imperfect.

Participants also rate the DEFault++ output positively. Most participants agree or strongly agree that the diagnosis is easy to understand, helps them decide on a repair, and would be useful in practice (\cref{fig:devstudy_results}).

\section{Threats to Validity}
\label{sec:defaultpp_threats}

\textbf{Threats to Internal Validity.} In our evaluation, information leakage could occur if runs from the same model--task pair were split across training and test folds. We address this risk with nested grouped cross-validation, which keeps each model--task pair within a single outer fold and limits preprocessing and hyperparameter selection to the inner training data. The joint training objective could also affect the evaluation because DEFault++ shares an encoder across diagnostic levels. As a result, the loss from one level may influence the representation used by another level. We mitigate the risk by restricting the separation loss to faulty samples within their ground-truth category and by evaluating its contribution through ablation.

\textbf{Threats to Construct Validity.} Mutation-generated faults may not fully represent faults that developers encounter in real-world transformer models~\citep{attention_taxonomy,deepcrime}. We reduce this risk by deriving each mutation operator from documented root causes in attention and DNN fault taxonomies~\citep{attention_taxonomy,faulttaxonomy,dlbugcharacterstics}, which are based on real-world examples. Class imbalance may also affect the evaluation results. We mitigate this risk by using macro-averaged F1 and inverse-frequency class weighting~\citep{buda2018systematic}. The mutation-killing test might fail to identify weak or noisy mutants, even when the intended change is successfully applied through fault injection. This may create a conservative fault set, but it improves label reliability because each retained faulty run differs statistically from its matched clean run. Extreme mutants may also inflate diagnostic performance when their effects are easy to detect~\citep{deepcrime}. We reduce this risk by varying fault severity and target layer, and by reporting mutation scores separately for each category. A final construct concern is the causal reading of the outputs. The training labels are causally grounded by construction, since each mutant differs from its clean counterpart through one targeted change under matched seeds. The Fault Propagation Graph and the diagnostic model, however, capture statistical associations rather than causal mechanisms, because the graph follows the computation graph and the model predicts mutation-generated labels. The RQ$_4$ group ablation and the real-world trace-to-code checks show that the reported feature groups are consistent with the predicted faults, but they do not establish a causal explanation, so we treat FPG edges as plausible propagation paths rather than as verified causal links.

\textbf{Threats to External Validity.} Generalization beyond the evaluated model--task pairs remains an external validity concern. We reduce this risk by using grouped outer folds, so each method is tested on pairs not seen during training or hyperparameter selection. Comparisons with prior techniques may also be affected by differences in prediction targets, since those techniques do not support DEFault++'s multi-level diagnoses (\cref{tab:related_work_comparison}). We therefore restrict direct baseline comparison to fault detection. In the developer study, differences between participants could affect repair choices. We reduce this risk with a counterbalanced within-subjects design, where each participant completes scenarios with and without DEFault++ support.

\section{Limitations and Future Work}
\label{sec:defaultpp_limitations}

DEFault++ currently diagnoses one injected fault per training run. This setting isolates each fault for categorization and root-cause diagnosis, but it does not cover programs with several co-occurring faults. Extending DEFault-bench to multi-fault configurations and studying how diagnostic behavior changes when several faults appear in the same run is a natural next step.

The Fault Propagation Graph (FPG) currently acts as a structural prior. Its edges mark possible propagation paths between transformer components, but they do not encode dependency strength or propagation mechanism. Calibrating these edges with fault-injection measurements and annotating them with dependency types would let DEFault++ distinguish stronger, weaker, and mechanism-specific propagation paths, and would let the model learn edge weights rather than treat every edge alike.

Our evaluation covers four encoder-only and three decoder-only models with 66M--125M parameters and 6--12 layers. It does not cover encoder-decoder models, vision models, multimodal models, mixture-of-experts architectures, or larger transformers. Since each mutation operator targets a component type, DEFault-bench extends to another model in the same architecture family by mapping its modules to those component types, although DEFault++'s trained classifier weights are model-specific and need retraining.

Our explanation evaluation measures faithfulness to DEFault++'s own decision process. DEFault-bench provides root-cause labels but not ground-truth feature-group explanations, so future work can strengthen the evaluation by adding independently validated feature-group annotations or by designing controlled faults with known feature-level effects.

Finally, our developer study is modest (N~=~21) and vignette-based rather than live, so we treat self-reported time as estimated effort rather than as a direct measure of task completion time. A larger live study that varies the type and severity of diagnostic errors would better characterize how practitioners use DEFault++ during debugging.

\section{Conclusion}
\label{sec:defaultpp_summary}
We present DEFault++, a transformer-specific hierarchical technique for fault detection, fault categorization, and root-cause diagnosis. DEFault++ extends our prior hierarchical DNN fault diagnosis to transformer architectures and reports which feature groups support each diagnosis. It organizes component-level measurements with a Fault Propagation Graph, a structural prior derived from the dependency paths in transformer computation. To train and evaluate it, we construct DEFault-bench, a benchmark of 5{,}556 labeled transformer runs generated with transformer-specific mutation testing. On DEFault-bench, DEFault++ reaches F1 of 0.826--0.909 for detection, 0.850--0.868 for categorization, and 0.851--0.867 for hierarchical root-cause diagnosis. In a developer study with 21 participants, repair accuracy rises from 57.1\% without DEFault++ to 83.3\% with it. The ablation and the developer study together point to one design lesson, that component-level measurements organized along the transformer's propagation structure separate faults that model-level training signals leave indistinguishable. We release DEFault-bench and the DEFault++ implementation in our replication package~\citep{defaultpp_repo}.

\bibliographystyle{ACM-Reference-Format}
\bibliography{refs}

\end{document}